\documentclass[11pt]{article}
\usepackage{comment}
\usepackage[margin=2.5cm]{geometry}
\usepackage{amsmath,amssymb,extarrows,mathtools,graphicx,subfigure,setspace,bbold}
\usepackage{cite}
\usepackage{slashed}
\usepackage[dvipsnames]{xcolor}
\usepackage{hyperref}
\hypersetup{colorlinks=true, linkcolor=blue, citecolor=magenta, linktoc=page}
\usepackage{tikz}
\usepackage{arydshln,leftidx,mathtools}
\usetikzlibrary{decorations.pathreplacing}
\numberwithin{equation}{section}
\newcommand{\be}{\begin{equation}}
\newcommand{\bea}{\begin{eqnarray}}
\newcommand{\eea}{\end{eqnarray}}
\newcommand{\ba}{\begin{align}}
\newcommand{\ea}{\end{align}}
\newcommand{\ee}{\end{equation}}

\usepackage{chngcntr}
\counterwithout*{figure}{section}

\begin{document}

\begin{titlepage}
\thispagestyle{empty}

\begin{flushright}
IPM/P-2017/012\\
\end{flushright}

\vspace{.4cm}
\begin{center}
\noindent{\Large \textbf{Entanglement in Lifshitz-type Quantum Field Theories}}\\

\vspace{2cm}

M. Reza Mohammadi Mozaffar and Ali Mollabashi
\vspace{1cm}

{\it School of Physics}
\\
{\it Institute for Research in Fundamental Sciences (IPM), Tehran, Iran}\\
\vspace{1cm}
Emails: {\tt m$_{-}$mohammadi, mollabashi@ipm.ir}

\vskip 2em
\end{center}

\vspace{.5cm}
\begin{abstract}
We study different aspects of quantum entanglement and its measures, including entanglement entropy in the vacuum state of a certain Lifshitz free scalar theory. We present simple intuitive arguments based on ``non-local" effects of this theory that the scaling of entanglement entropy depends on the dynamical exponent as a characteristic parameter of the theory. The scaling is such that in the massless theory for small entangling regions it leads to area law in the Lorentzian limit and volume law in the $z\to\infty$ limit. We present strong numerical evidences in $(1+1)$ and $(2+1)$-dimensions in support of this behavior. In $(2+1)$-dimensions we also study some shape dependent aspects of entanglement. We argue that in the massless limit corner contributions are no more additive for large enough dynamical exponent due to non-local effects of Lifshitz theories. We also comment on possible holographic duals of such theories based on the sign of tripartite information.
\end{abstract}

\end{titlepage}

\newpage

\tableofcontents
\noindent
\hrulefill

\onehalfspacing

\section{Introduction}
Studying the physics of non local correlations due to the quantum entanglement in quantum many-body systems, quantum field theories and especially holographic field theories has gained increasing attention during the last decade
 \cite{Casini:2009sr, Calabrese:2009qy, Laflorencie:2015eck, Rangamani:2016dms}. Furthermore, in order to quantify entanglement and have a deeper understanding of it, different measures has been studied so far such as entanglement entropy (EE), mutual information and logarithmic negativity \cite{Casini:2008wt, Calabrese:2009ez, Calabrese:2012ew, DeNobili:2016nmj}. In particular EE is a good measure of entanglement for pure stets, since it behaves as an entanglement monotone decreasing under LOCC \cite{Vidal:2000}.

The recipe for computing EE is simple and straightforward. Considering the simplest setup where the physical system consists of two subsystems such that $\mathcal{H}_{\rm tot.}=\mathcal{H}_A\otimes\mathcal{H}_{\bar{A}}$. Integrating out the degrees of freedom which live in the second subsystem, i.e., $\bar{A}$, one can find a reduced density matrix for $A$ which we denote by $\rho_A$. EE is given by the corresponding von Nuemann entropy for $\rho_A$. In contrast to the thermal entropy, the entanglement entropy is nonvanishing at zero temperature and thus it is a good probe for studying properties of quantum phase transitions. Also it captures some characteristic properties of the underlying theory which the best known examples are even dimensional conformal field theories (see \cite{somereviews} for reviews). 

At a quantum critical point, a physical system typically exhibits a Lifshitz scaling symmetry as follows\cite{Lifshitz,Hertz:1976zz}
\bea\label{Lifshitzscaling}
t\rightarrow \lambda^z t,\;\;\;\;\;\;\;\vec{x}\rightarrow \lambda \vec{x},
\eea
where $z$ denotes the dynamical critical exponent. The above inhomogeneous scaling in general breaks down the beloved Lorentz invariance. Merely in the very special case of $z=1$, the relativistic dynamics is recovered.
A huge amount of efforts has been made to understand the entanglement structure in relativistic QFTs, which has resulted in many interesting features and of course many open questions. This constructs the very most literature of quantum entanglement in the context of QFTs.
Although there have been some attempts so far to analyse the entanglement structure in the non-relativistic cases with $z\neq 1$ (see \cite{Fradkin:2006mb,Solodukhin:2009sk,Zhou:2016ykv} on the QFT side and  \cite{Alishahiha:2012cm,Alishahiha:2014cwa,Fonda:2014ula,Ecker:2015kna,Kusuki:2017jxh,Hosseini:2015gua} in the context of gauge/gravity duality\footnote{QFTs with Lifshitz symmetry has been widely studied in the context of gauge/gravity duality, see e.g. \cite{Kachru:2008yh, Alishahiha:2012nm, Mozaffar:2012bp, Taylor:2015glc}.}), still there are several open questions about the role of $z$ in the entanglement structure of such theories.

Here one of our main motivations is to construct a more stringent analysis of the nature of quantum correlations at Lifshitz critical points. An interesting tool to do so is studying quantum entanglement in theories with Lifshitz symmetry. We will consider the simplest example of such QFTs, which is a free scalar theory with generic dynamical exponent  $z$. This family of QFTs represents several interesting features which is not of our interest here (see e.g. \cite{Alexandre:2011kr} for a review).
A similar analysis has been previously done for the same theory but for $0<z<1$ in \cite{Nezhadhaghighi:2012vz, Nezhadhaghighi:2013mba, Nezhadhaghighi:2014pwa, Rajabpour:2014osa}. We show that the behavior of $z>1$ is totally different from what was found for $z<1$.\footnote{We would like to thank Tadashi Takayanagi for bringing our attention to these references during the final steps of this work.} 

The first step to study the entanglement structure of these theories is to focus on well-known measures such as entanglement entropy. This is often a formidable task to carry out analytically in the context of QFTs. In principle there are different strategies including, replica trick\cite{Calabrese:2009qy}, numerical approaches or even `generalized' holographic prescriptions \cite{Ryu:2006bv} which one can employ them to overcome this problem and gain some information about the entanglement structure of theories with Lifshitz symmetry.

In this paper we will tackle the problem numerically using the so called correlator method \cite{Peschel2003, Peschel2009}.  This method is based on a semi-analytic approach leading to the reduced density matrix (as well as its $n$-th power) corresponding to a given subregion. The correlator method which is applicable to quadratic field theories is based on reducing all correlation functions to two-point functions restricted to a given subregion. This method which we will shortly review in the next section makes it possible to calculate entanglement and Renyi entropies in free field theories. We will employ this method to study entanglement measures in the ground state of a free scalar theory with Lifshitz symmetry.  

\begin{figure}
\centering
\includegraphics[scale=.28]{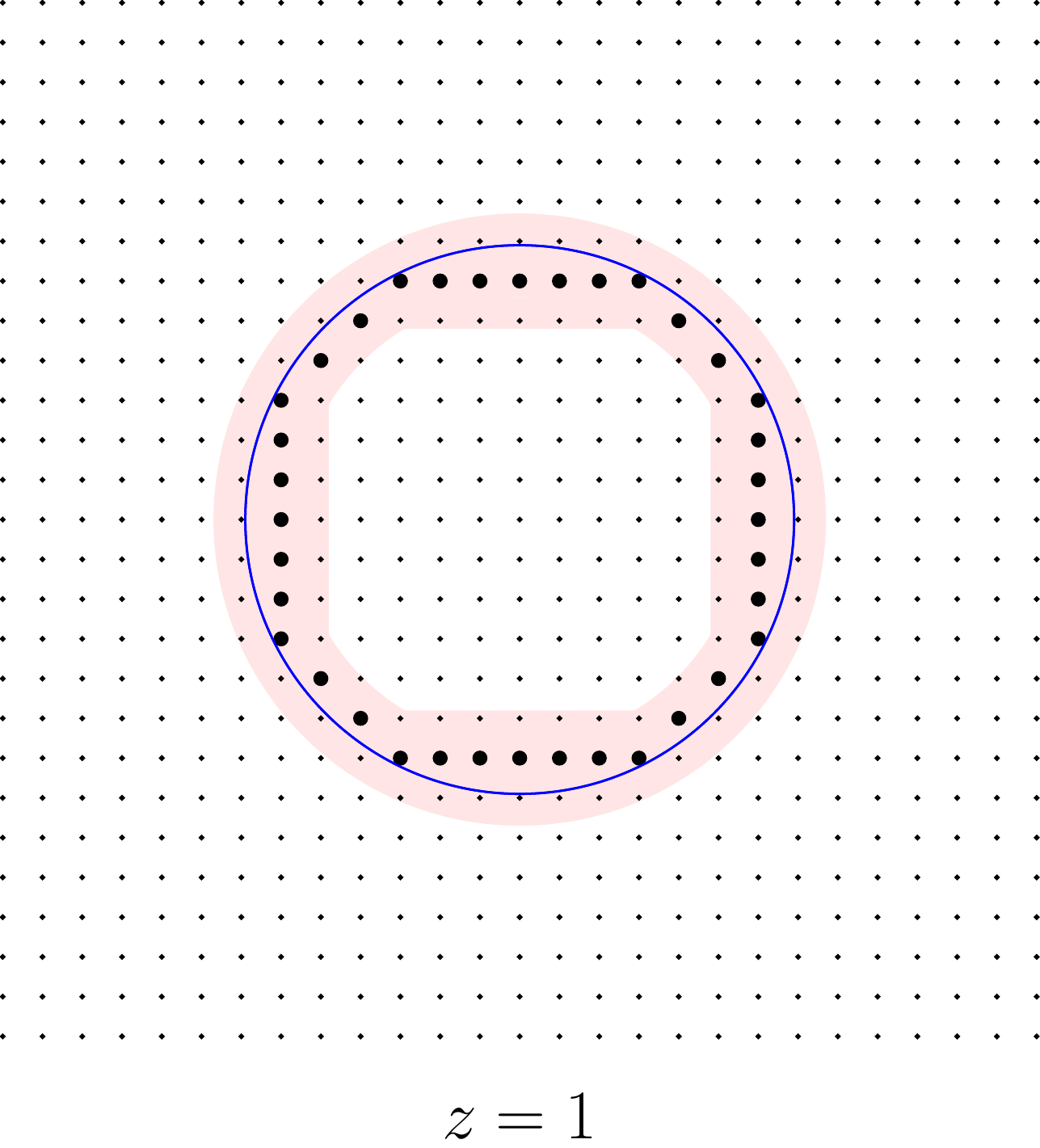}
\hspace{1mm}
\includegraphics[scale=.28]{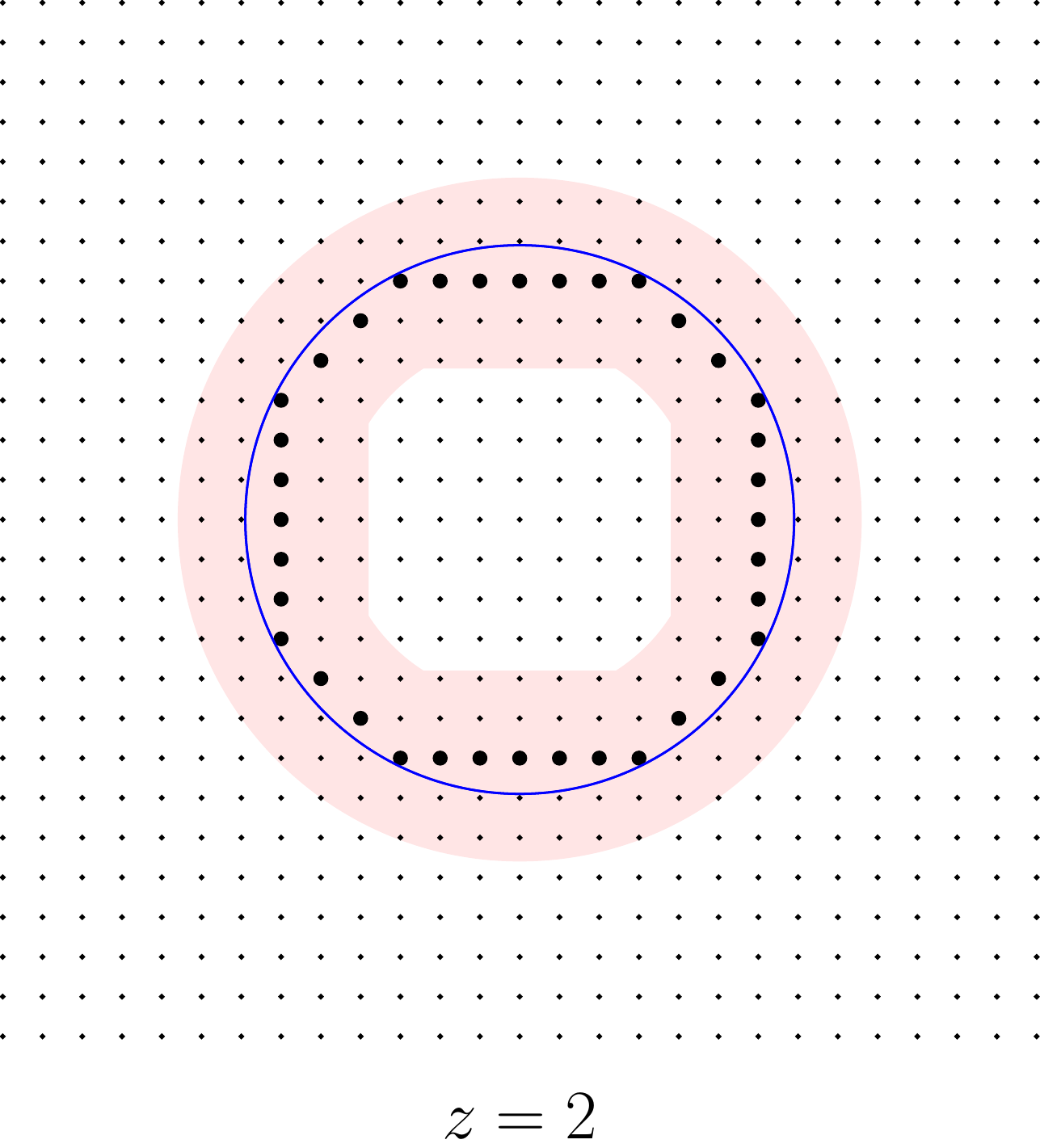}
\hspace{1mm}
\includegraphics[scale=.28]{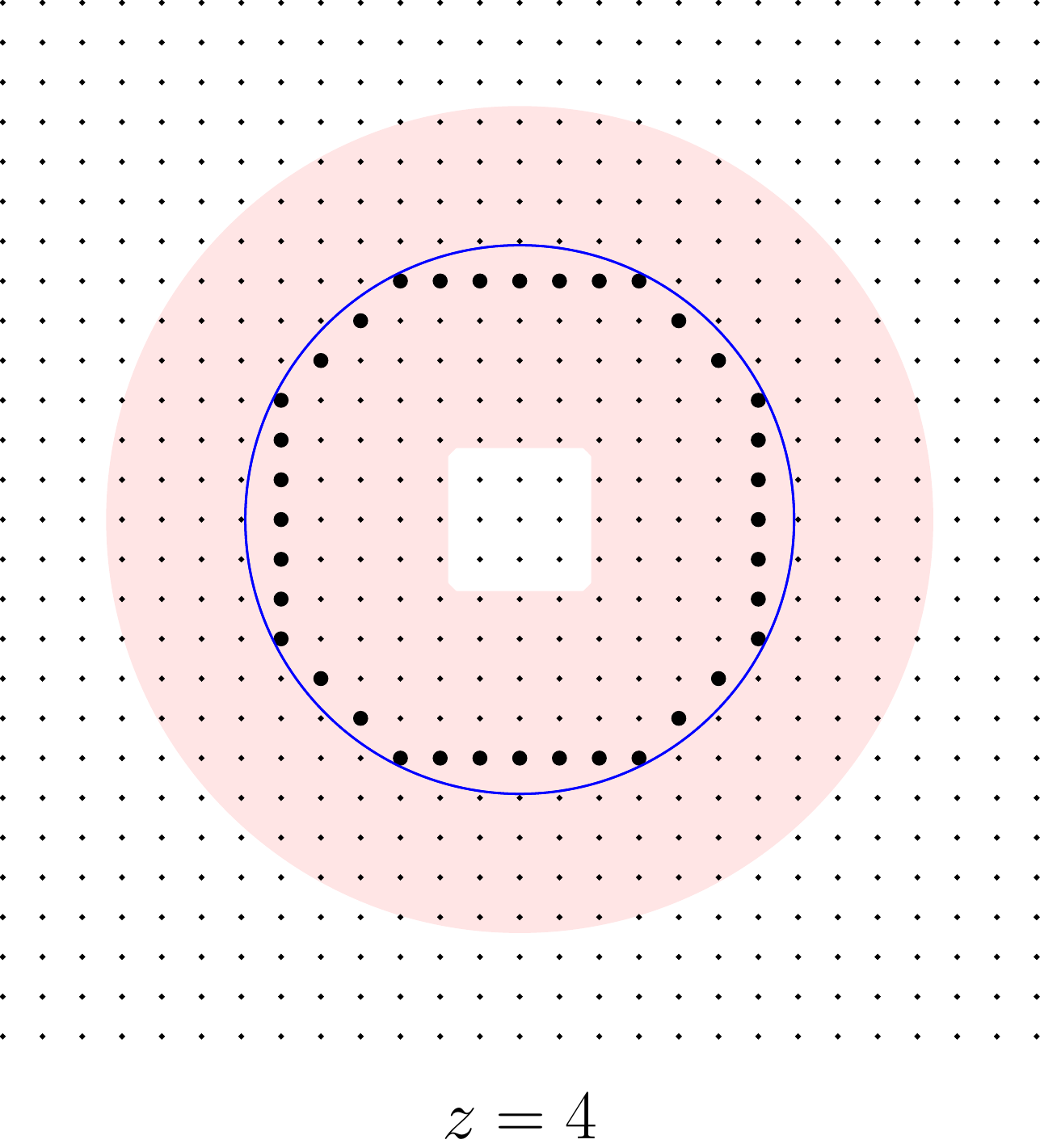}
\hspace{1mm}
\includegraphics[scale=.28]{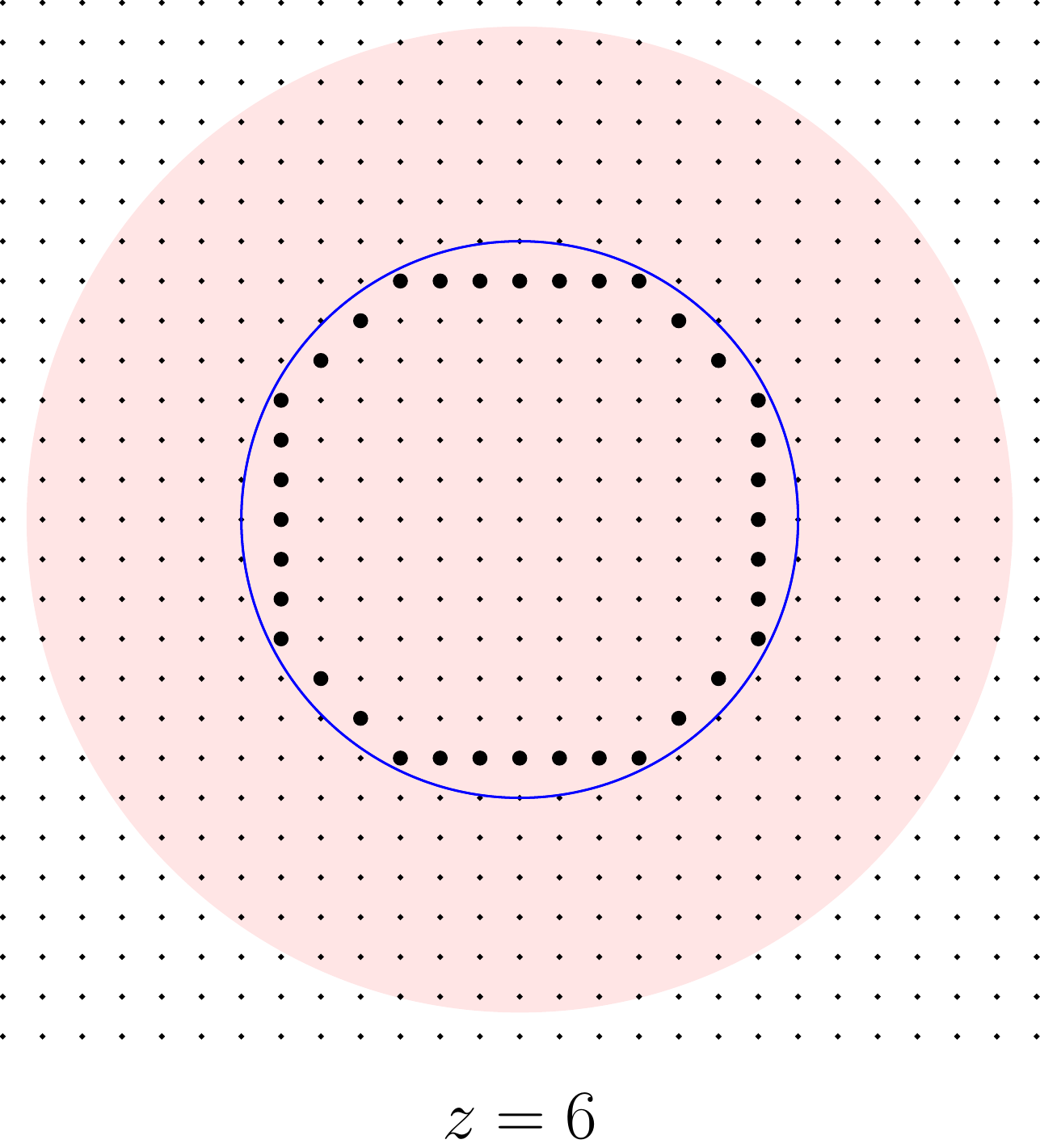}
\caption{The nearest points inside a disk entangling region to its boundary are shown in green. Those points which are correlated to the green ones due to the discrete kinetic term of this theory are shown with a red shadow. The left panel belongs to the Lorentzian case $z=1$. This shows how we intuitively understand the area law of entanglement for regions which their characteristic length, say $\ell$, is much bigger than the lattice spacing (the inverse of the UV cut-off). Moving from the left panel to the right, the dynamical exponent is increasing. One can see that for a disk with a radius $\sim 3.5$ in units of lattice spacing, for $z> 6$, since all points inside the entangling region are correlated with the green ones, we expect entanglement measures to scale with the volume (instead of the area) of the entangling region. We will show numerically that this is the correct scaling of entanglement entropy for large enough $z$ in the following of this paper. Although this figure belongs to $(2+1)$-dimensions, its horizontal (vertical) slices describe what happens in $(1+1)$-dimensions and it can be generalized to higher dimensions straightforwardly.}
\label{fig:L-NL2p1}
\end{figure}

We will focus on free massless and massive scalar theory with dynamical exponent $z\geq 1$. In order to perform numerical calculations we put the theory on a square lattice. The main difference between theories with Lifshitz symmetry and their Loretzian counterpart, i.e. the same theory with the dynamical exponent $z=1$, is that the number of lattice sites which are `correlated' together due to the spatial part of the kinetic term is $z$-dependent. To be more precise, for a positive integer $z$, the kinetic term has a spatial derivative of order $2z$, thus it correlates $(2z+1)$ lattice points together. This fact revises the well-known intuitive understanding of the area law of entanglement in local field theories which is based on the short range correlation due to the kinetic term. In Fig.\ref{fig:L-NL2p1} we have illustrated the intuitive counterpart in this case that which the area law transits to a volume law while the dynamical exponent increases.  

The rest of this paper is organized as follows: in section \ref{sec:Review} we briefly introduce our QFT setup and correlator method which we use in our analysis. We report our results for different entanglement measures of our Lifshitz theory of interest in the massless regime in section \ref{sec:1p1} and show that EE obeys a volume law for large $z$. We also study the shape dependence of EE in $2+1$ dimensions and report our results for different entangling regions such as a disk and square and present strong evidences for emergent volume law behavior of EE at large $z$. We also show that corner contributions to EE get away of being local effects in such theories for $z>1$. In section \ref{sec:massive} we continue our study by considering a massive scalar theory in different dimensions. The last section is devoted to the conclusions and several possible interesting directions to investigate in future works.

\textbf{Note added:}
Reference \cite{He:2017wla} which appeared after our paper has some overlap with our study in (1+1)-dimensions. The results are compatible where overlap exists.  

\section{Correlator Method in Lifshitz-type QFTs}\label{sec:Review}
\subsection{QFTs with Lifshitz Symmetry}
The starting point of our analysis is the following nonrelativistic action for a free massive scalar field in $(d+1)$-dimensions\cite{Alexandre:2011kr}\footnote{Note that the most general theory with such a symmetry may also include  terms with lower order spatial derivatives, say $\phi\left(\partial\cdot\partial\right)^\nu\phi$ with $\nu$ being an integer and $\nu<z$. These terms have generic coefficients in the action which (for simplicity) we have set all of them to vanish in our analysis (see \cite{Alexandre:2011kr} for detailed analysis of such theories).}
\bea\label{action}
I=\frac{1}{2}\int dt d\vec{x} \left[\dot{\phi}^2-\sum_{i=1}^{d}(\partial_i^z \phi)^2-m^{2z} \phi^2\right],
\eea
which in the massless case is invariant under the scaling introduced in Eq.\eqref{Lifshitzscaling}. According to this action the mass dimensions are given as follows
\bea	
[t]=-z,\;\;\;[\vec{x}]=-1,\;\;\;[m]=1,\;\;\;[\phi]=\frac{d-z}{2},
\eea
where $z$ should be an integer parameter. In the following discussion we always consider $z>1$ case.

In this theory in the massless case one can easily find the scaling behavior of the ground state two point correlator to be  \cite{Rajabpour:2014osa}
\be
\langle\phi(0)\phi(r)\rangle=\int \frac{d^dk}{(2\pi)^d}\frac{e^{i\vec{k}\cdot\vec{x}}}{2|k|^z}\sim r^{-d+z}.
\ee
This behavior shows that the power law growth of correlator as the dynamical exponent increases. This is in agreement with what we have explained intuitively in Fig.\ref{fig:L-NL2p1}. 

Using the momentum density conjugate to $\phi$, i.e., $\pi=\frac{\partial \mathcal{L}}{\partial \dot{\phi}}$, the corresponding expression for the Hamiltonian density can be written
\bea\label{hamil}
\mathcal{H}=\frac{1}{2}\left(\pi^2+\sum_{i=1}^d(\partial_i^z \phi)^2+m^{2z}\phi^2\right).
\eea
In the following we concentrate on case $d=1$ having in mind that the generalization to arbitrary $d$ is straightforward. In order to push our calculations we introduce a UV cut-off and thus put the theory on a lattice. The Hamiltonian density on a square lattice is given by
\bea\label{lathamil}
\mathcal{H}=\frac{1}{2}\sum_{n=0}^{N-1}\left[\pi_n^2+\left(\sum_{k=0}^{z}(-1)^{z+k} {{z}\choose{k}} \phi_{n-1+k}\right)^2+m^{2z}\phi_n^2\right],
\eea
where we consider a lattice with $N$ sites and without loss of generality we set the lattice spacing equal to unity. Note that $\phi_n$ and $\pi_n$ satisfy standard commutation relations, i.e., $[\phi_n,\phi_m]=[\pi_n,\pi_m]=0$ and $[\phi_n,\pi_m]=i\delta_{nm}$. In the following sections we will consider periodic and Dirichlet boundary conditions which correspond to $\phi_N=\phi_0 (\pi_N=\pi_0)$ and $\phi_N=\phi_0=0 (\pi_N=\pi_0=0)$ respectively. 

In the case of periodic boundary conditions, using the translation invariance one can diagonalize the Hamiltonian given in Eq.\eqref{lathamil} with the following Fourier transformations
\bea
\phi_n=\frac{1}{\sqrt{N}}\sum_{k=0}^{N-1}\tilde{\phi}_k e^{\frac{2\pi i k n}{N}},\;\;\;\;\pi_n=\frac{1}{\sqrt{N}}\sum_{k=0}^{N-1}\tilde{\pi}_k e^{\frac{2\pi i k n}{N}}.
\eea
Expanding these new canonical variables in terms of creation and annihilation operators, i.e., 
\bea
\tilde{\phi}_k=\frac{1}{\sqrt{2\Omega_k}}(a_k+a_{-k}^\dagger),\;\;\;\tilde{\pi}_k=i\sqrt{\frac{\Omega_k}{2}}(a_{-k}^\dagger-a_k),
\eea
the Hamiltonian density becomes
\bea
\mathcal{H}=\sum_{k=0}^{N-1}\Omega_k(a_{k}^\dagger a_{k}+\frac{1}{2}),
\eea
where 
\bea\label{Omega}
\Omega^2_k=m^{2z}+\left(2\sin\frac{\pi k}{N}\right)^{2z}.
\eea
We have used the following sum rule
\bea
4\sum_{j=0}^{z-1}(-1)^{j} {{2z}\choose{z-1-j}}\sin^2(j+1)\frac{\pi k}{N}=\left(2\sin\frac{\pi k}{N}\right)^{2z}.
\eea
to derive the dispersion relation given  in Eq.\eqref{Omega}.
Notice that although in Eq.\eqref{action} we consider $z$ as an integer parameter, actually Eq.\eqref{Omega} shows the exact analytic continuation to non integer dynamic critical exponents. 
Employing the above definitions of dynamical fields in terms of creation and annihilation operators the vacuum correlators are given as follows
\bea\label{corr1}
\langle \phi_i \phi_j \rangle = \frac{1}{N}\sum_{k=0}^{N-1}\frac{1}{2\Omega_k}e^{2\pi i \frac{k}{N}(i-j)},\;\;\;\;
\langle \pi_i \pi_j \rangle = \frac{1}{N}\sum_{k=0}^{N-1}\frac{\Omega_k}{2} e^{2\pi i \frac{k}{N}(i-j)},
\eea
where for the infinite lattice limit, i.e., $N\rightarrow \infty$, becomes
\bea\label{corr2}
\langle \phi_i \phi_j \rangle = \int_0^1 dx\, \frac{1}{2\Omega(x)}e^{2\pi i x(i-j)},\;\;\;
\langle \pi_i \pi_j \rangle = \int_0^1 dx\, \frac{\Omega(x)}{2}e^{2\pi i x(i-j)},
\eea
where $\Omega(x)^2=m^{2z}+(2\sin\pi x)^{2z}$. Of course the correlators only depend on $(m-n)$ due to the translation invariance of the system.\footnote{See \cite{Casini:2005zv} for similar analysis of Lorentzian free scalar theory.} Note that according to the above expressions, in the case of periodic boundary condition the massless limit, i.e., $m\rightarrow 0$, is not well-defined due to the existence of a zero mode. This makes us consider a non-zero mass in order to regularize the corresponding divergences.

One way to get rid of the zero mode is to break the translational symmetry of the system. As an example one can consider Dirichlet boundary condition on the borders of the total system. Since there is no translation invariance in this system, one should consider a Fourier sine transformation for the fields which gives \cite{Calabrese:2012nk}, 
\bea
\phi_n=\sqrt{\frac{2}{N}}\sum_{k=1}^{N-1}\tilde{\phi}_k \sin \frac{\pi k n}{N},\;\;\;\;\pi_n=\sqrt{\frac{2}{N}}\sum_{k=1}^{N-1}\tilde{\pi}_k \sin \frac{\pi k n}{N}.
\eea
The analysis then proceeds in analogy with the case of periodic boundary condition. Evaluating the corresponding two point functions, we find that
\bea\label{corrDir}
\langle \phi_i \phi_j\rangle = \frac{1}{N}\sum_{k=1}^{N-1}\frac{1}{\tilde{\Omega}_k}\sin \frac{\pi k i}{N}\sin \frac{\pi k j}{N},\;\;\;\;
\langle \pi_i \pi_j \rangle = \frac{1}{N}\sum_{k=1}^{N-1}\tilde{\Omega}_k\sin \frac{\pi k i}{N}\sin \frac{\pi k j}{N},
\eea
where $\tilde{\Omega}_k=\Omega_{\frac{k}{2}}$. In this case the zero mode is no more excited and there is no divergence in Eq.\eqref{corrDir} while $m\to 0$. In this case we will consider the massless limit\footnote{An explicit analysis of zero modes in EE in relativistic free scalar theory has been recently carried out in \cite{Yazdi:2016cxn}.}. 

\subsection{Review of Correlator Method}
As we mentioned in the introduction section, in order to find the entanglement entropy, one should construct the reduced density matrix corresponding to the subregion of interest. Since we are interested in a free (quadratic) theory, we use the correlator method first introduced in \cite{Peschel2003}\footnote{See also \cite{Casini:2009sr} for a pedagogical presentation of this method.}.

This method is based on a very simple idea: If we are with the expectation value of any operator restricted to region $A$, which we denote by $\langle\mathcal{O}_A\rangle$, in principle this would fix the reduced density matrix corresponding to  this subregion uniquely through
\be\label{eq:OA}
\mathrm{Tr}\left[\rho_A\mathcal{O}_A\right]=\langle\mathcal{O}_A\rangle.
\ee
In the context of quadratic theories, thanks to Wick's theorem one can reduce the right hand side of Eq.\eqref{eq:OA} to all possible 2-point functions of the theory. In this way it is not hard to show that the eigenvalues of the reduced density matrix which has the following form
\be
\rho_A=\mathcal{N}\exp\left({\sum_i\varepsilon_ia_i^\dagger a_i}\right)
\ee
(with $\mathcal{N}$ a normalization constant) is given by 
\bea
\varepsilon_k=2\coth^{-1}2\nu_k,
\eea
where $\nu_k$'s are the eigenvalues of $C=\sqrt{X.P}$ which itself is given by the following 2-point functions 
\be
X_{ij}=\langle \phi_i \phi_j \rangle\;\;\;\;\;,\;\;\;\;\;P_{ij}=\langle \pi_i \pi_j \rangle.
\ee
Using the definition of entanglement and Renyi entropies it is a very simple task to show that 
\begin{align}
S_A&=\sum_{k=1}^{n_A}\left[\left(\nu_k+\frac{1}{2}\right)\log\left(\nu_k+\frac{1}{2}\right)-\left(\nu_k-\frac{1}{2}\right)\log\left(\nu_k-\frac{1}{2}\right)\right],\label{EE}\\
S^{(n)}_A&=\frac{1}{n-1}\sum_{k=1}^{n_A}\log\left[\left(\nu_k+\frac{1}{2}\right)^n-\left(\nu_k-\frac{1}{2}\right)^n\right],
\end{align}
where $n_A$ is the number of lattice points enclosed by the entangling region $A$. 
The method just outlined is very general and can be employed to any quadratic QFT with different matter content such as fermions in any space-time dimensions. Also note that using this method one can find other entanglement measures, e.g., mutual and tripartite information as we will do in the subsequent sections.

\begin{center}
\begin{figure}
\includegraphics[scale=0.43]{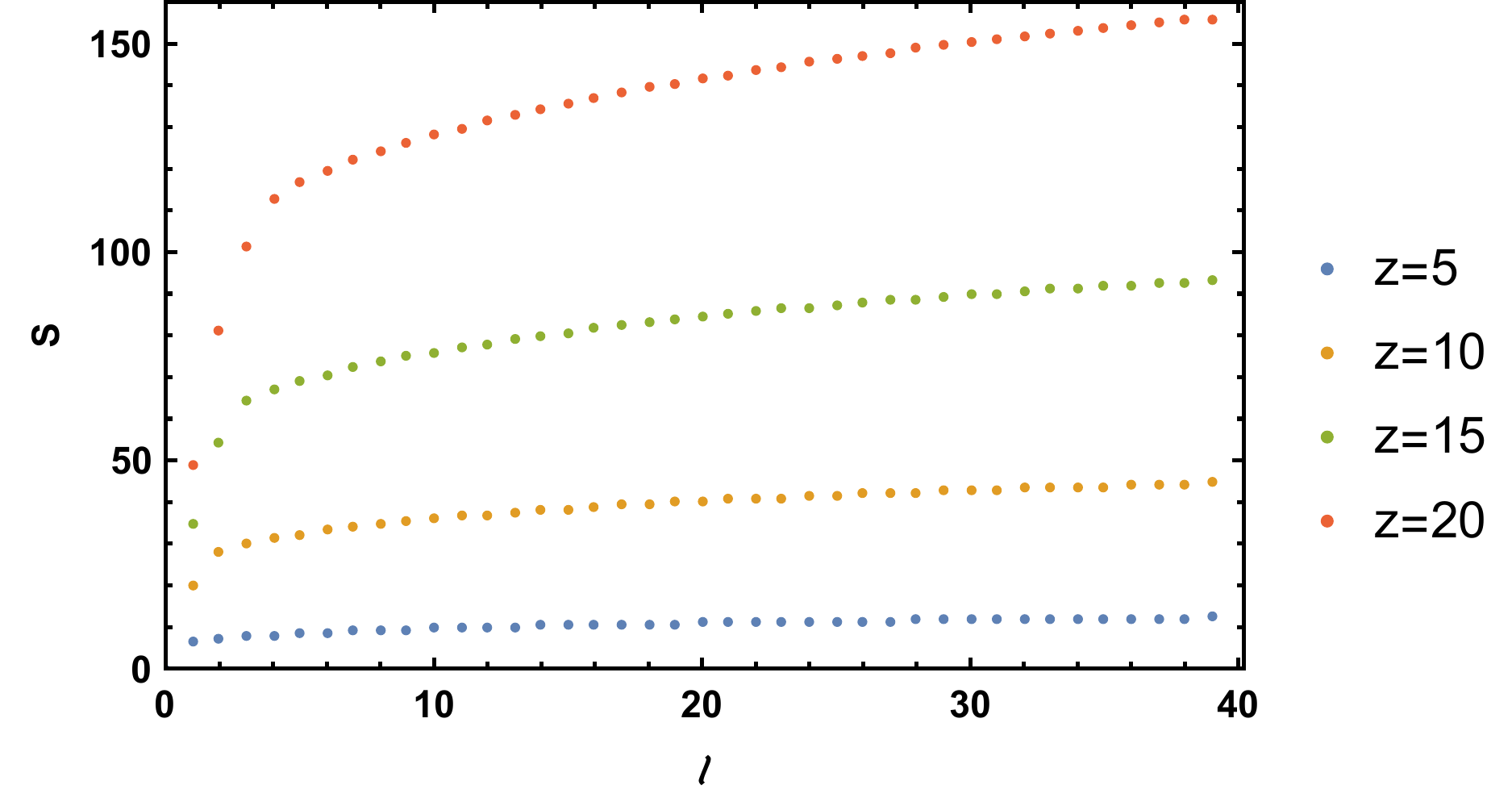}
\includegraphics[scale=0.75]{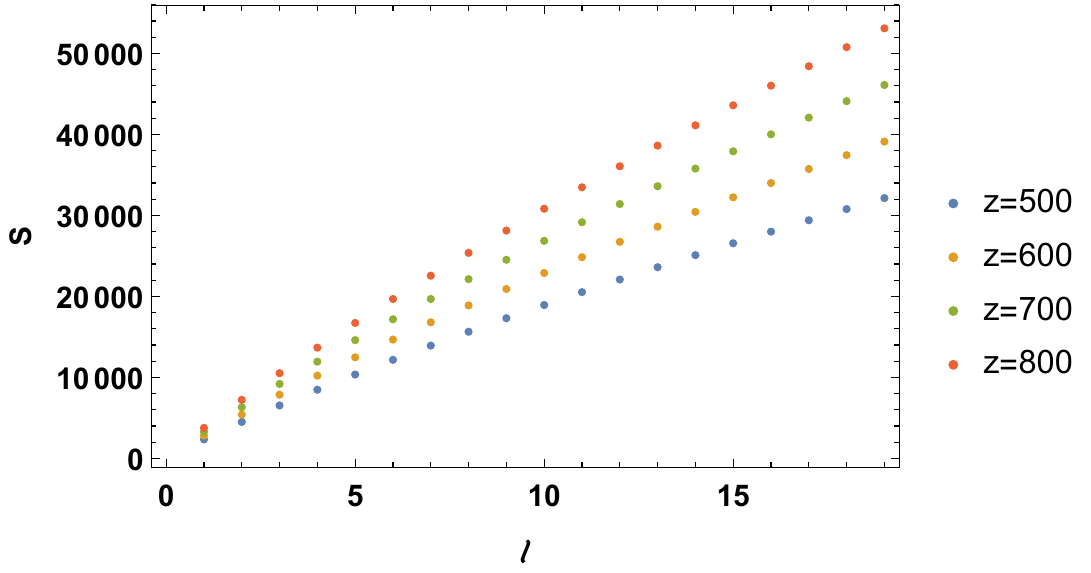}
\caption{EE as a function of $\ell$ for different values of $z$. Here we consider Dirichlet boundary condition for a massless scalar with $N=500$ (left) and $N=20000$ (right).}
\label{fig:EE-scalar-Dir}
\end{figure}
\end{center}
\section{Entanglement Measures in Lifshitz-type QFTs}\label{sec:1p1}
\subsection{Entanglement Measures in Massless Scalar Theory}
\subsubsection{($1+1$)-dimensions}
In this section we study the entanglement and Renyi entropies in massless Lifshitz field theory introduced in Eq.\eqref{action}. As we have mentioned before, the existence of zero modes restricts us to impose Dirichlet boundary condition in which the zero modes are not excited and we do not need an IR cutoff.

We consider the discrete version on a lattice and use Eq.\eqref{corrDir} to find the corresponding correlators. The results for entanglement entropy are collected in Fig.\ref{fig:EE-scalar-Dir} where we demonstrate numerical results for different values of dynamical exponent. Based on this figure, it is evident that the value of entanglement entropy increases while the dynamical exponent is increased. Such a behavior is not surprising because as we have seen in the discrete version of the Hamiltonian in Eq.\eqref{lathamil}, for larger dynamical exponents the number of lattice points which are coupled together increases and thus the correlation between points inside and outside the entangling region increases. Indeed, for generic $z(>1)$, the number of correlated points due to the kinetic term increases as $2z+1$ (this is illustrated in any horizontal or vertical slice of different panels of Fig.\ref{fig:L-NL2p1}).

\begin{center}
\begin{figure}
\includegraphics[scale=0.78]{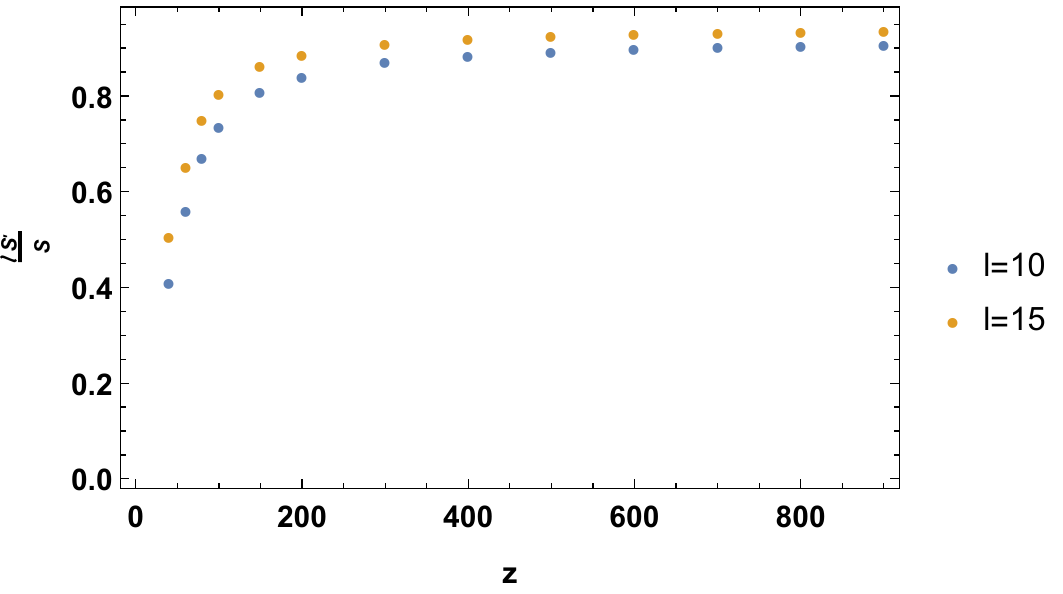}
\includegraphics[scale=0.78]{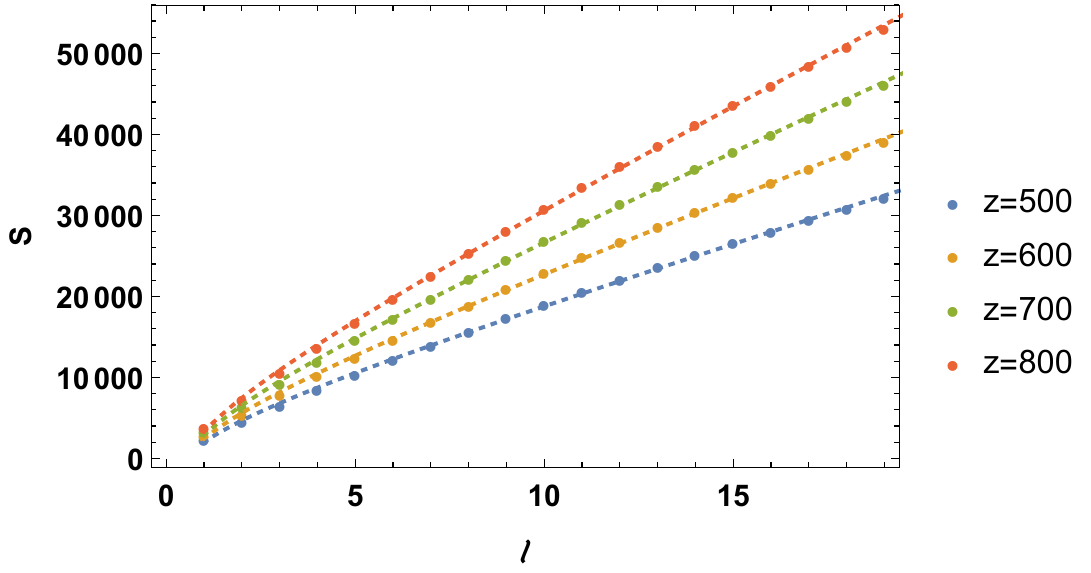}
\caption{\textit{Left}: $\frac{\ell S'}{S}$ as a function of $z$ for different values of $\ell$. For large values of $z$ this quantity approaches to a constant value exhibiting volume law behavior. \textit{right}: Fitting data for EE as a function of $\ell$ according to Eq.\eqref{fitfunction} for different values of $z$. Here we consider Dirichlet boundary condition for a massless scalar with $N=20000$.}
\label{fig:check}
\end{figure}
\end{center}

Interestingly, according to Fig.\ref{fig:EE-scalar-Dir} for large values of $z$ and small enough entangling regions the theory exhibits a volume law scaling of the entanglement entropy. By small enough we mean $\ell/N\ll 1$. This is in contrast with the behavior of entanglement entropy in typical vacuum states of a local relativistic Hamiltonian which exhibits area law scaling and in $(1+1)$-dimensions is replaced with a logarithmic scaling. Again, from the expression of lattice Hamiltonian, i.e., Eq.\eqref{lathamil}, it is clear that by increasing the critical exponent the theory starts to show nonlocal effects such that for $z\gg 1$ the corresponding QFT becomes highly nonlocal. This is why we expect and numerically verify the volume law rather than the area law. Actually according to Fig.\ref{fig:EE-scalar-Dir} one expects that EE is proportional to $\ell$ with
a certain power, i.e., $S\sim \ell^\alpha$. But in order to confirm the volume law, the asymptotic
behavior should be $\alpha|_{z\rightarrow \infty}\rightarrow 1$. In the left panel of Fig.\ref{fig:check} we check this asymptotic behavior considering $\frac{\ell S'}{S}$ as a function of the dynamical exponent for two different entangling regions. According to these curves, $\frac{\ell S'}{S}$ approaches to unity while $z$ goes to infinity which confirms the volume law.

\begin{figure}
\begin{center}
\includegraphics[scale=0.7]{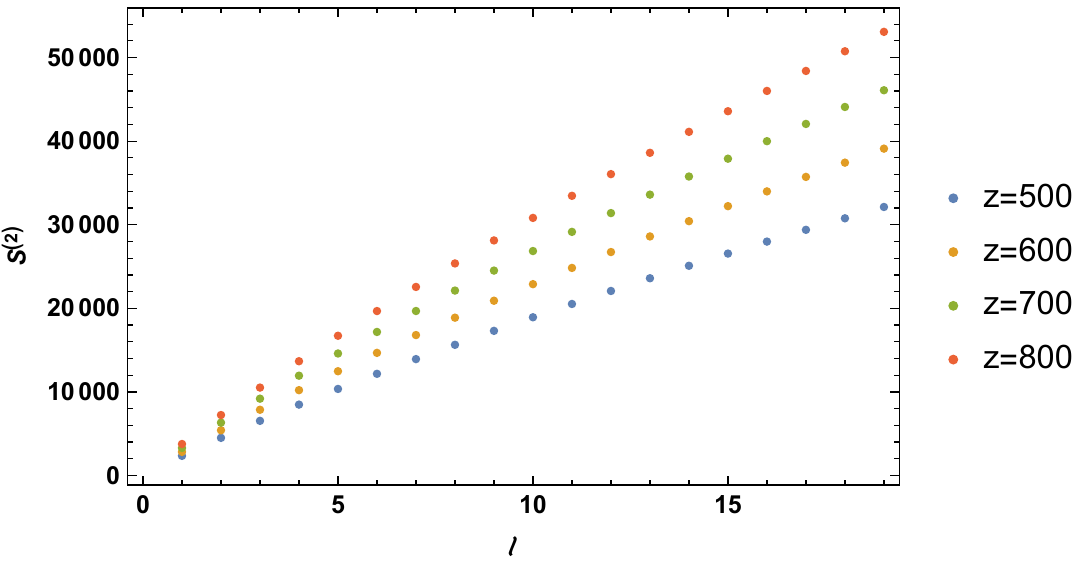}
\includegraphics[scale=0.7]{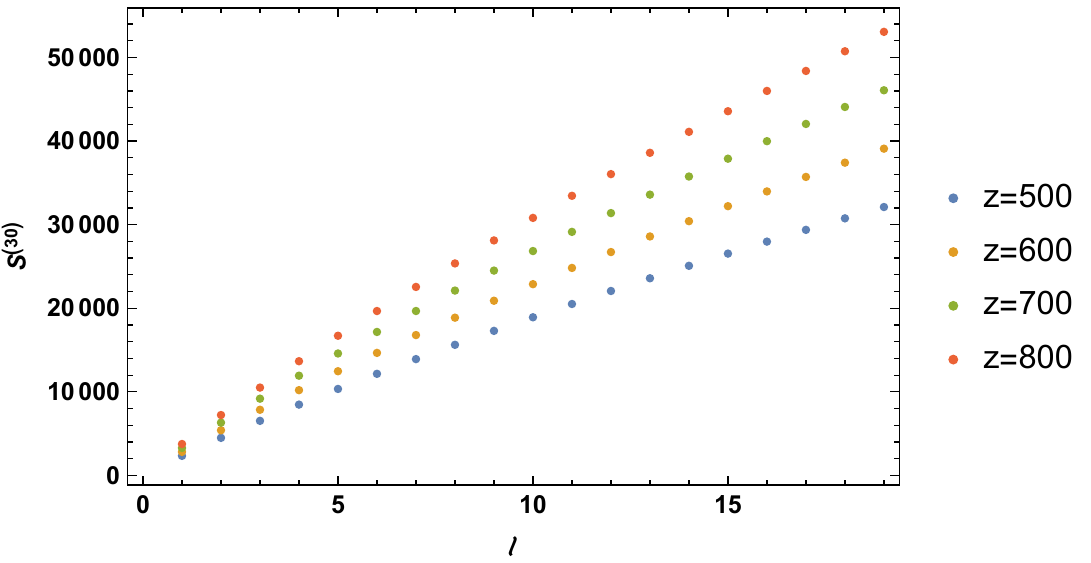}
\end{center}
\caption{Renyi entropy as a function of $\ell$ for large values of $z$ with $m=0$ and $N=20000$.
}
\label{fig:RE-scalar-Dir}
\end{figure}

Volume law scaling of entanglement entropy has been previously observed in several field theories with nonlocal effects \cite{Barbon:2008ut, Fischler:2013gsa, Karczmarek:2013xxa, Karczmarek:2013jca, Shiba:2013jja} (See also \cite{Rabideau:2015via} for a perturbative study of entanglement entropy in nonlocal theories.).\footnote{Volume law behavior of entanglement entropy also appears in lattice models with local Hamiltonian but due to broken translational invariance \cite{Vitagliano:2010db} or even for non-geometric entanglement entropy such as field space entanglement \cite{Mollabashi:2014qfa, Mozaffar:2015bda}.}
The model of our interest for large dynamical exponents is similar to a certain non-local scalar theory studied in \cite{Shiba:2013jja}. In Fig.\ref{fig:RE-scalar-Dir} we have shown that this scaling behavior also happens for Renyi entropies  for large values of $z$. 

We would like to clarify a very important point about theories which show a volume scaling of entanglement entropy in a certain range of parameters. As long as we are dealing with pure states, one may worry about some fundamental features of entanglement entropy in these theories. Explicitly one may ask about the validity of $S_A=S_{\bar{A}}$ ($\bar{A}$ is the complement of $A$) is such theories for regions which are small enough to show a volume law. Lets consider $A$ the small region which its entropy scales with its volume. In this case the complement region $\bar{A}$ is a large subregion which dose not have a volume law. If we have the wrong impression that the $S_{\bar{A}}$ scales with its area, then we will lose $S_A=S_{\bar{A}}$! The important point is that $S_{\bar{A}}$ does not scale with the area of $\bar{A}$ in this case. This is simply because the boundary of $A$ and $\bar{A}$ are the same ($\partial A=\partial \bar{A}$) and the characteristic length of the boundary is smaller than the non-locality scale. In this case it is not enough to consider a strip of dof's around $\partial \bar{A}$ to find the leading term of entanglement entropy. In other words for regions which are large such that their complements show a volume law, the entanglement entropy does not scale with area but with the volume of their complement. With this in mind $S_A=S_{\bar{A}}$ is still valid in these theories.\footnote{We would like to thank Sergey Solodukhin, Charles Rabideau and Noburo Shiba for useful discussions about this point.}

For small values of the dynamical exponent we still expect the logarithmic scaling to be the leading contribution to entanglement entropy. For a fixed subregion as the dynamical exponent increases we expect the scaling to tend to volume law. In order to investigate this competition between the contributions due to logarithmic and volume scaling more precisely, we study the ratio of these portions with considering the following fit function
\bea\label{fitfunction}
S^{(z)}(\ell) = \mathcal{S}^{(z)}_{\rm vol.} \frac{\ell}{\epsilon}+\mathcal{S}^{(z)}_{\rm univ.} \log \frac{\ell}{\epsilon}+ S^{(z)}_0,
\eea
where $S^{(z)}_0$ is a constant (and also non-universal) term where we drop it in the following discussion by defining a subtracted EE as $\Delta S^{(z)}(\ell)\equiv S^{(z)}(\ell)-S^z_0$.
Analysing the numerical data shows that the ratio of the volume term with respect to the subtracted entanglement entropy, i.e., $\frac{\mathcal{S}^{(z)}_{\rm vol.}}{\Delta S^{(z)}}$ should tend to $\ell/\epsilon$ in the large $z$ limit. This behavior shows that for large values of critical exponents the volume law contribution to EE substantially is dominant. Based on our numerical results, we propose that there should be a leading piece in $S^{(z)}(\ell)$, whose dependence on $\ell$ characterizes a transition from area to volume law at a Lifshitz quantum critical point, and its form for $d=1$ is given by
\bea\label{eq:s1p1}
S^{(z)}(\ell)= \# \left(\frac{\ell}{\epsilon}\right)^{1-\frac{1}{z}}+\cdots,
\eea
where for $z=1$ should be replaced by a logarithmic scaling. In the right panel of Fig.\ref{fig:check} we demonstrate the numerical results which are in agreement with our proposal. Of course the scaling introduced in Eq.\eqref{eq:s1p1} is not the most general form but rather the simplest function supporting such a transition between area and volume law. We expect whatever the exact analytic behavior is, which can be found via different methods, e.g., working out the propagator of scalar field on an $n$-sheeted plane (with Lifshitz symmetry), having the same behavior as Eq.\eqref{eq:s1p1} in the $\epsilon\to0$ limit.

\begin{figure}
\begin{center}
\includegraphics[scale=0.8]{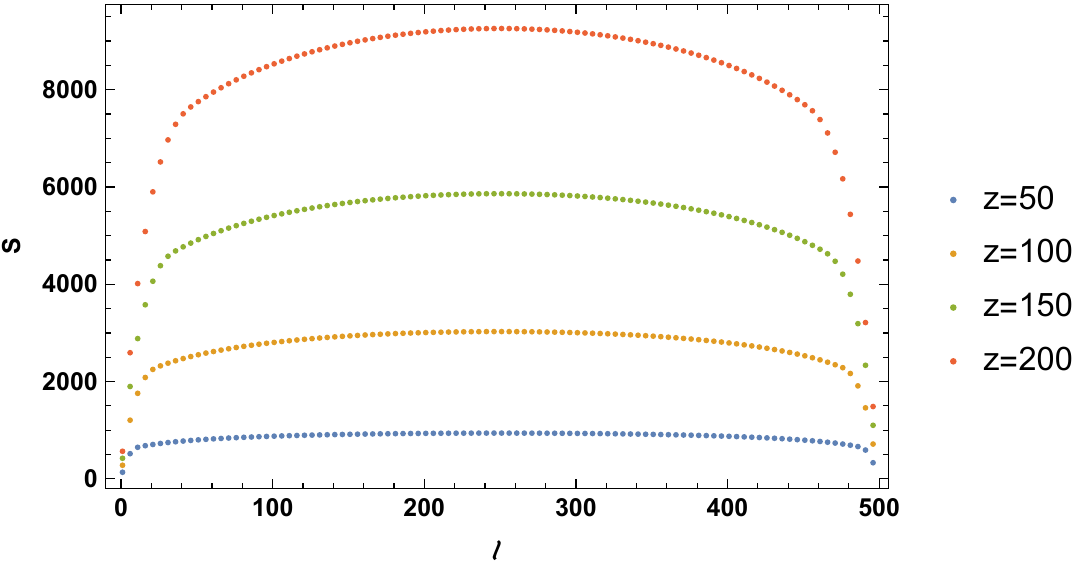}
\end{center}
\caption{EE for large regions $\ell \sim N$ showing $S_A=S_{\bar{A}}$.}
\label{fig:SASB}
\end{figure}

It is also worth to note that although for large values of $z$ and for $\ell \ll N$, the scaling of EE differs from the usual area law, but for pure state that we have considered the $S_A=S_{\bar{A}}$ condition still holds. We demonstrate this behavior in Fig.\ref{fig:SASB}.

\begin{figure}
\begin{center}
\includegraphics[scale=0.4]{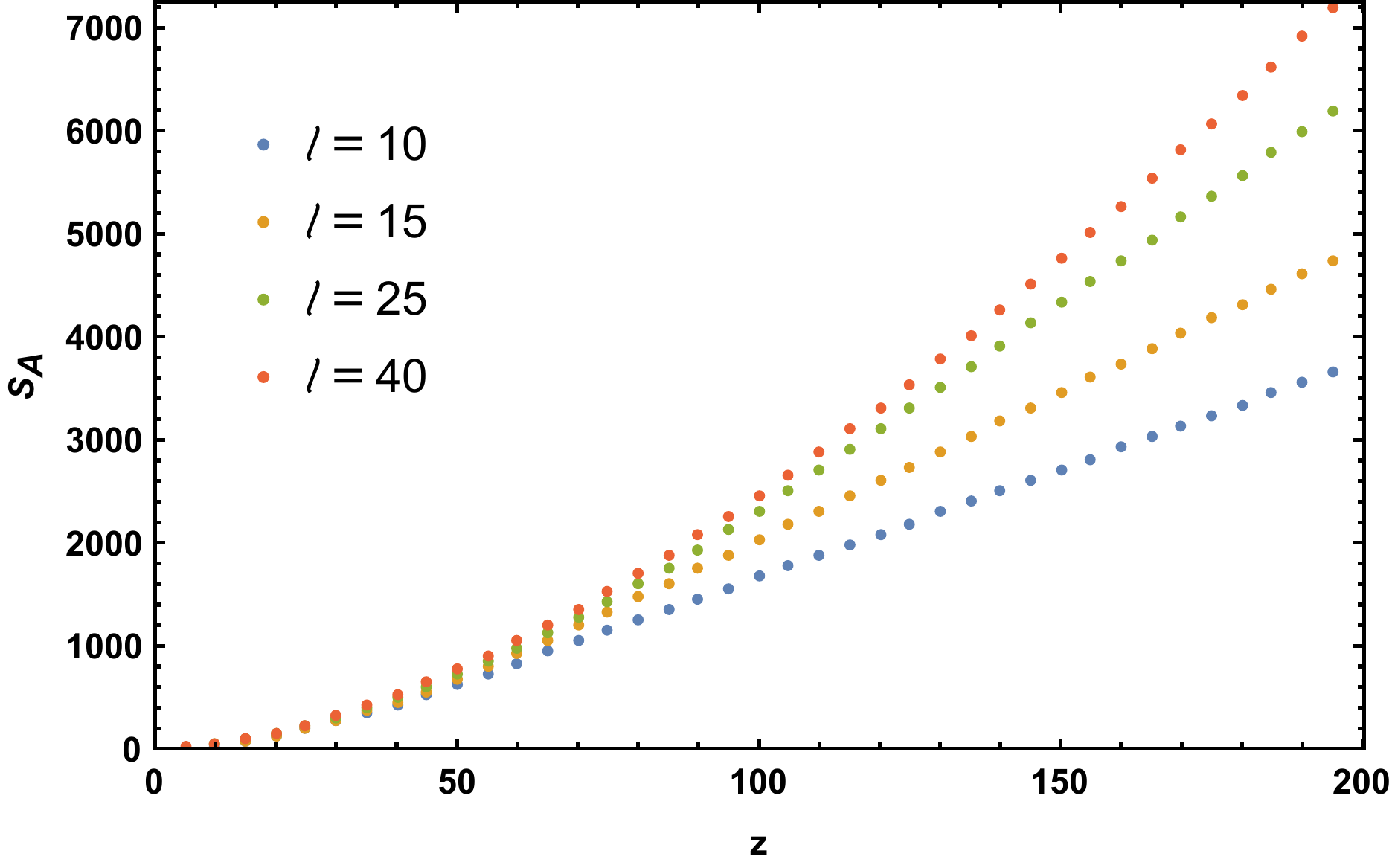}
\end{center}
\caption{EE as a function of $z$ for entangling regions with different lengths. Here we have set $N=500$.}
\label{fig:Sofz1}
\end{figure}

A relevant question would be how the entropy depends on the dynamical exponent $z$. We have also studied this numerically which the result is shown in Fig. \ref{fig:Sofz1}. As one can see there is a quadratic growth as $z$ starts to increase from the Lorentzian case. The validity of this regime depends on the size of the entangling region. In the large $z$ limit the numerical results shows linear growth of entanglement entropy as a function of $z$. The smaller the region is, the faster the entropy growth enters the linear regime. This is in agreement with the analytic result found in \cite{He:2017wla} for a different type of entangling regions called $p$-alternating sublattice.   

\subsection*{Mutual and Tripartite Information}
After analyzing the behavior of entanglement entropy at a critical point with Lifshitz scaling, we now proceed with considering other entanglement measures. In particular we are interested in mutual and tripartite information between different subregions.\footnote{Some aspects of mutual information for $z<1$ have been previously studied in \cite{Nezhadhaghighi:2013mba}.}
These quantities are defined in terms of EE as follows
\begin{align*}
I(A,B)&=S(A)+S(B)-S(A\cup B),\\
I^{[3]}(A,B,C)&=S(A)+S(B)+S(C)-S(A\cup B)-S(A\cup C)-S(B\cup C)+S(A\cup B\cup C),
\end{align*}
where $A$, $B$ and $C$ are three entangling regions with length $a$, $b$ and $c$ respectively which we consider in the following configuration:
\begin{figure}[h!]
\begin{center}
\includegraphics[scale=0.7]{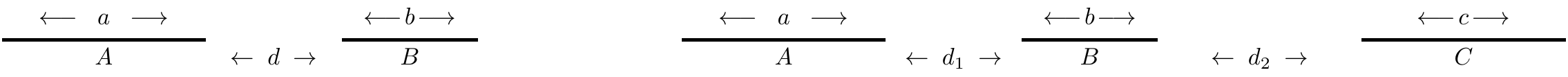}
\end{center}
\end{figure}

Although the mutual information is always positive, which is a reminiscent of subadditivity of EE, the tripartite information can be positive, negative or zero depending on the quantum field theory and subregion configuration of interest. It is important to mention that in spite of EE which is divergent, these two quantities are defined such that the contributions of different subregions to UV divergent part cancel each other and the resultant expression becomes finite. Using the above definition for mutual and tripartite information and Eq.\eqref{EE} one can find these quantities for a Lifshitz QFT. 

\begin{figure}
\begin{center}
\includegraphics[scale=.8]{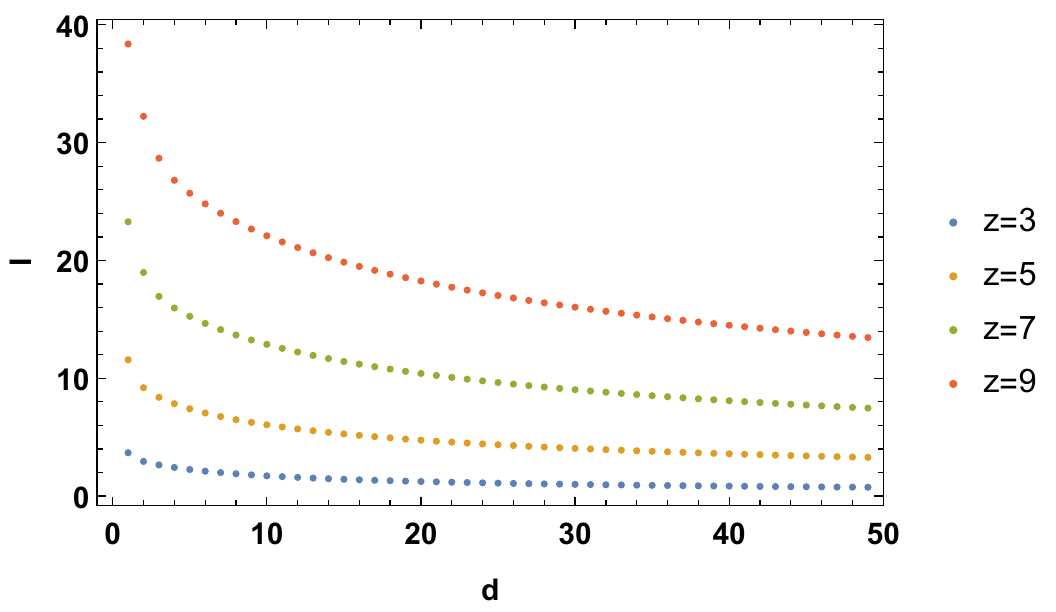}
\end{center}
\caption{Mutual information as a function of the separation between subregions for different values of dynamical exponent for $a=b=20$, $m=0$ and $N=500$.}
\label{fig:I2I3}
\end{figure}

Let us first focus on mutual information. Here for simplicity we restrict our analysis to a configuration where the length of regions $A$ and $B$ are equal i.e. $a=b$. The generalization to more generic configurations is straightforward. For such a configuration, we expect the mutual information between two fixed subregions to be a decreasing function of the separation between them which we denote by $d$. The results are collected in Fig.\ref{fig:I2I3}.
Based on this figure there are two points in hand: First, as the dynamical exponent increases, the mutual information between subregions for a fixed configuration increases. This in agreement with what we explained about the increase of EE as the dynamical exponent increases. The second point is related to the well-known fact that mutual information between fixed subregions vanishes while the separations between subregions increases. Our results shows that the critical separation between subregions increases with the dynamical exponent.\footnote{To be more concrete, we should emphasis that the location of this vanishing point depends on the ratio $\frac{d}{\ell}$ where $\ell$ is the length of the each subregion and $d$ is the separation between them.} Once again, this behavior is expected due to the enhancement of spatial correlations for larger values of $z$.

The behavior of tripartite information is qualitatively similar to mutual information. One can check that similar to what happens in Lorentzian free scalar theory, tripartite information is always positive in this Lifshitz scalar theory. This is why we think that our results may not be comparable to some results available in the literature which are found by means of Ryu-Takayanagi proposal on geometries with asymptotic Lifshitz symmetry (see for instance \cite{Alishahiha:2012cm, Kusuki:2017jxh}). It is worth to remind the reader that in the context of holographic QFTs, it is a well-known fact that holographic tripartite information, at least for states which have a classical gravity dual, must be always negative. The proof of this statement is based on the minimality condition in Ryu-Takayanagi formula \cite{Hayden:2011ag}.\footnote{A generalisation for other holographic entanglement measures in some specific limits has be done in \cite{Alishahiha:2014jxa}.}

\subsubsection{$(2+1)$-dimensions}
\begin{figure}
\centering
\includegraphics[scale=.3]{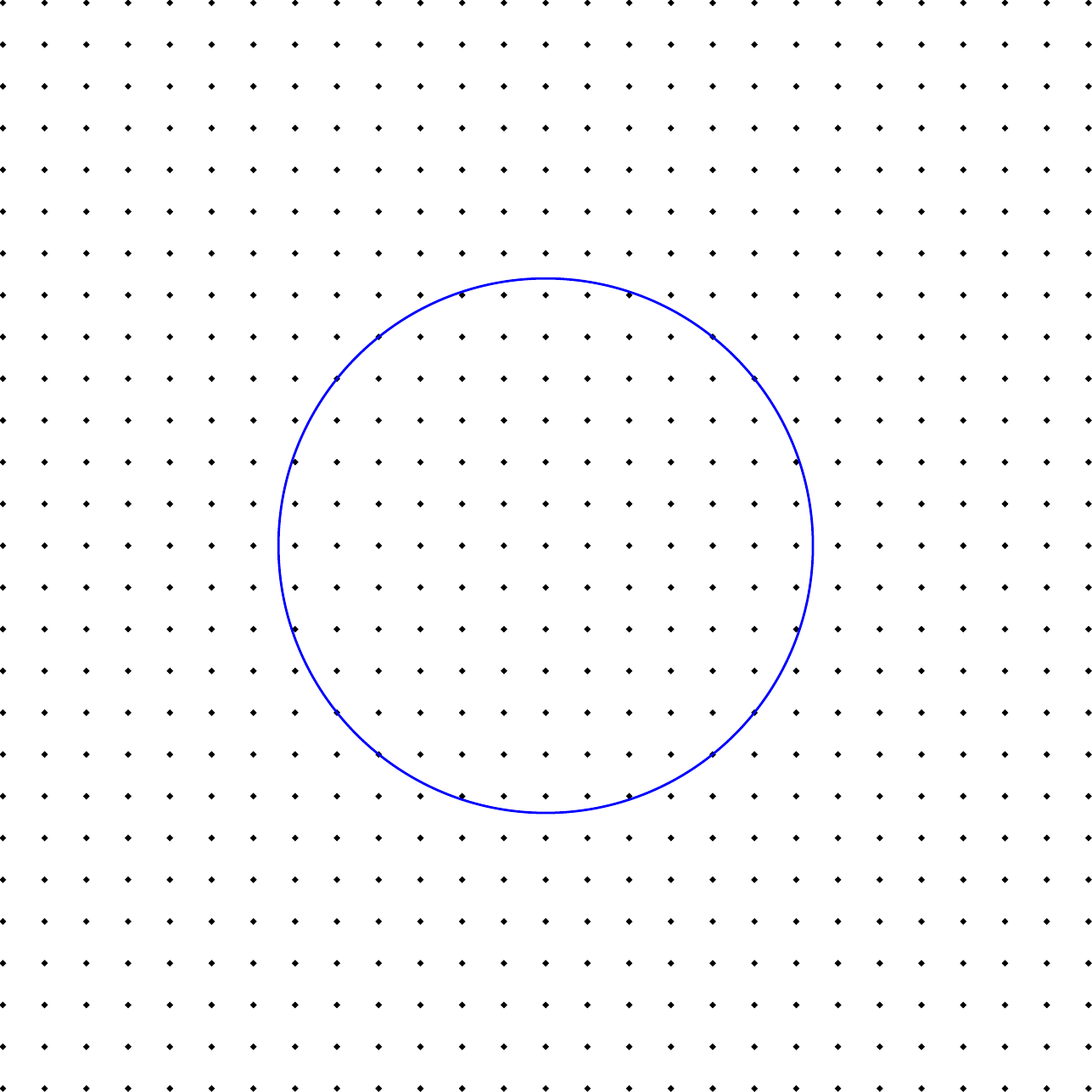}
\hspace{2mm}
\includegraphics[scale=.3]{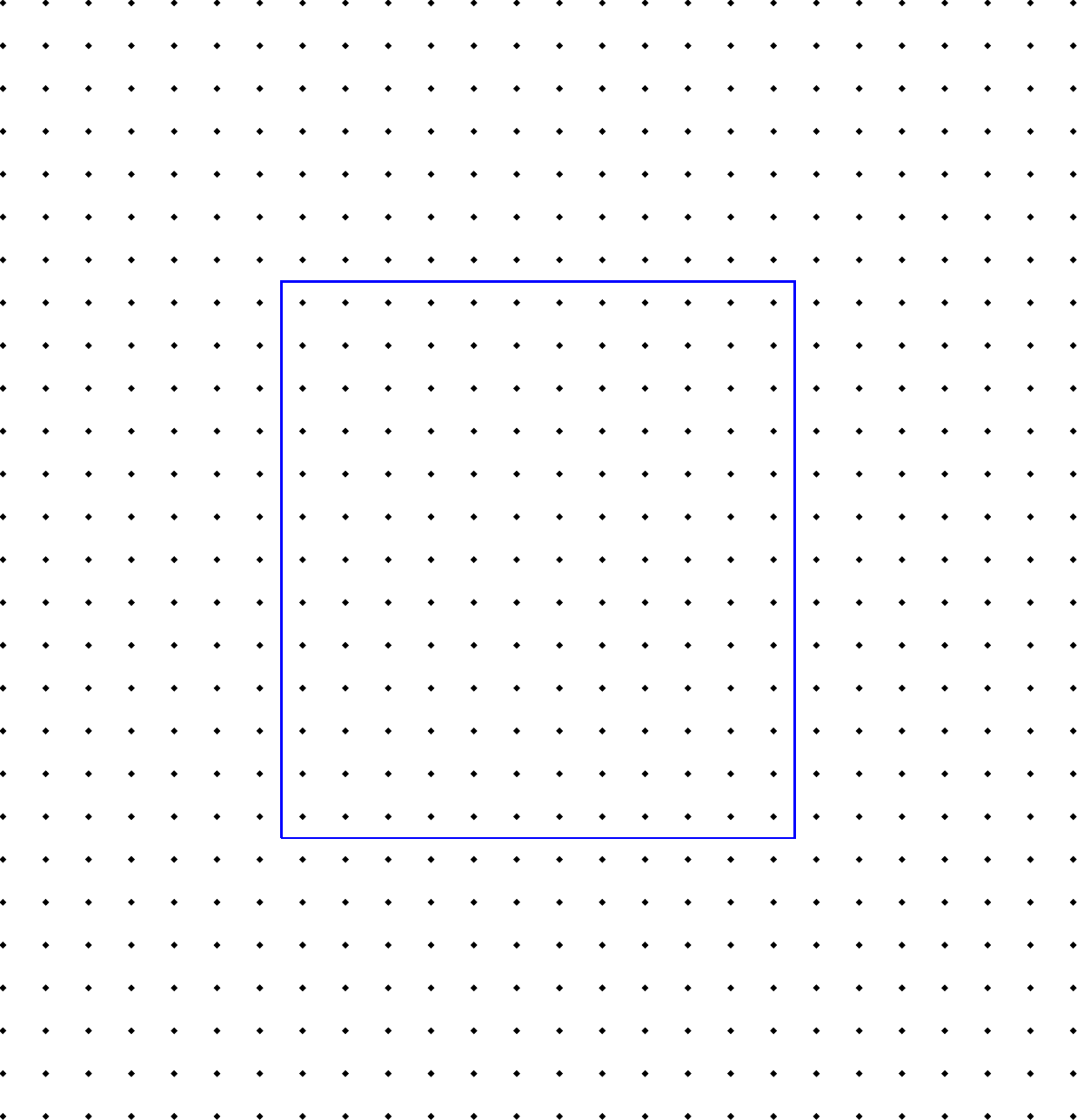}
\hspace{2mm}
\includegraphics[scale=.3]{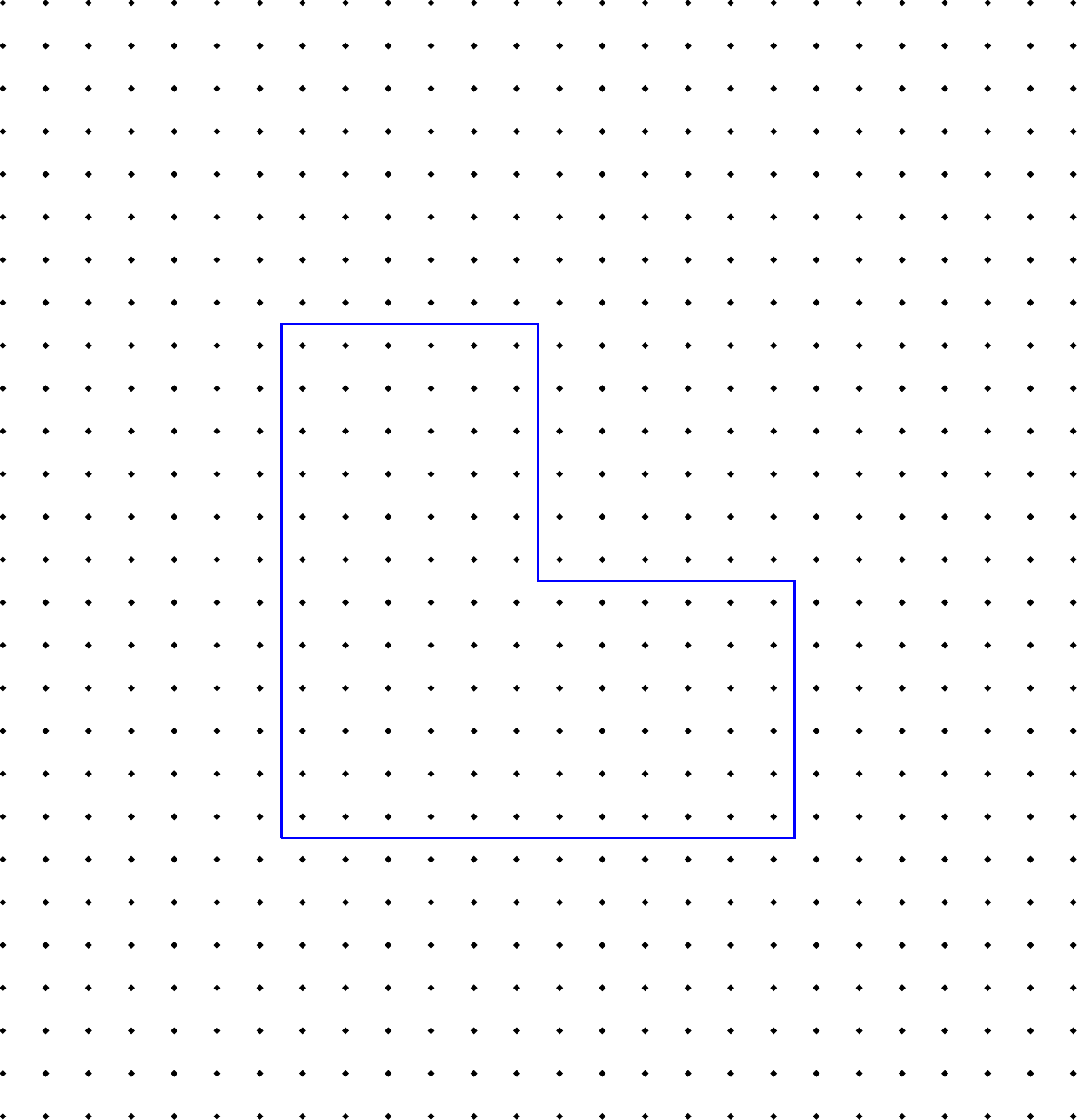}
\hspace{2mm}
\includegraphics[scale=.3]{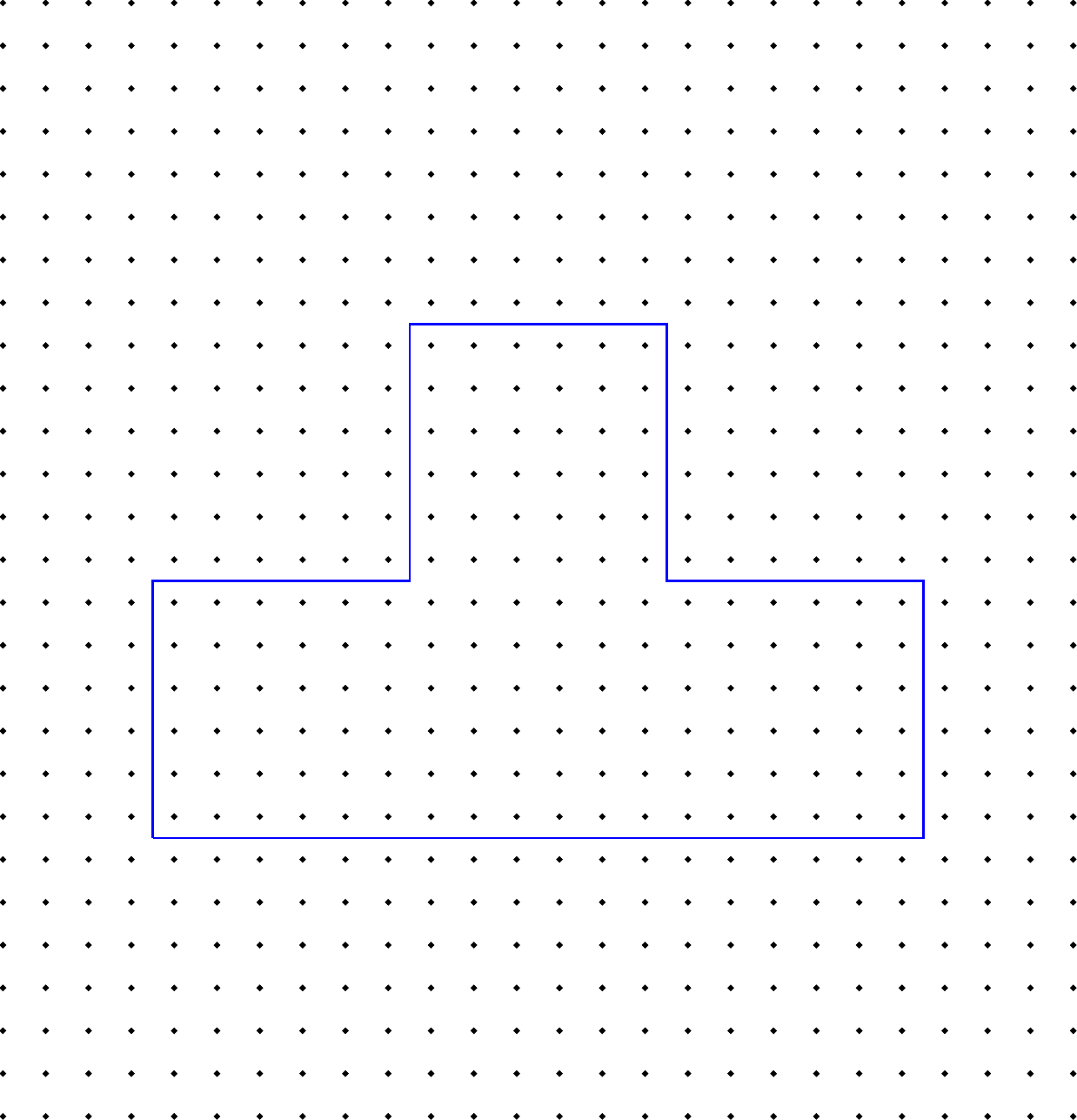}
\caption{Different entangling regions on a constant time slice of a $(2+1)$-dimensional Lifshitz field theory. From left the first panel shows a disk entangling region and the second one is a square region. The third and forth shapes are deformations of the square region. The area of the shape in the third panel is equal to that of the square but they have different number of corners (6 and 4 respectively). The shape in the forth panel with 8 corners, has the same volume as the square. We call the third one `a-shape' and the forth one `v-shape'.}
\label{fig:shapes}
\end{figure}

In this section we proceed our previous analysis in one higher dimension which enables us to more investigate the emergence of volume law and also shape dependence of EE in presence of dynamical exponent $z$.\footnote{Entanglement entropy for Lorentzian free scalar theory in $(2+1)$-dimensions has been studied in \cite{Casini:2006hu, Casini:2009sr}.} In order to do so we first generalize our method which was introduced in section \ref{sec:Review} to $(2+1)$-dimensions. One can straightforwardly show that the vacuum correlators with periodic boundary conditions in this case are given by the following expressions
\begin{align}
\begin{split}
\langle \phi_{i,j} \phi_{r,s} \rangle &= \frac{1}{2N_xN_y}\sum_{k_x=0}^{N_x-1}\sum_{k_y=0}^{N_y-1}\frac{1}{\Omega_{k_x,k_y}}e^{2\pi i \frac{k_x}{N_x}(i-r)}e^{2\pi i \frac{k_y}{N_y}(j-s)},\\
\langle \pi_{i,j} \pi_{r,s} \rangle &= \frac{1}{2N_xN_y}\sum_{k_x=0}^{N_x-1}\sum_{k_y=0}^{N_y-1}\Omega_{k_x,k_y}e^{2\pi i \frac{k_x}{N_x}(i-r)}e^{2\pi i \frac{k_y}{N_y}(j-s)},
\end{split}
\end{align}    
where
$$\Omega^2_{k_x,k_y}=m^{2z}+\left(2\sin\frac{\pi k_x}{N_x}\right)^{2z}+\left(2\sin\frac{\pi k_y}{N_y}\right)^{2z},$$
$(N_x,N_y)$ and $(k_x,k_y)$ are the number of lattice sites and the momentum in our two spatial directions.
It is also straightforward to generalize the vacuum correlators for Dirichlet boundary condition in this case. This is given by Fourier double sine transformation which gives
\begin{align}
\begin{split}
\langle \phi_{i,j} \phi_{r,s} \rangle &= \frac{2}{N_xN_y}\sum_{k_x=1}^{N_x-1}\sum_{k_y=1}^{N_y-1}\frac{1}{\tilde{\Omega}_{k_x,k_y}}\sin\frac{\pi k_x i}{N_x}\sin\frac{\pi k_y j}{N_y}\sin\frac{\pi k_x r}{N_x}\sin\frac{\pi k_y s}{N_y},\\
\langle \pi_{i,j} \pi_{r,s} \rangle &=\frac{2}{N_xN_y}\sum_{k_x=1}^{N_x-1}\sum_{k_y=1}^{N_y-1}\tilde{\Omega}_{k_x,k_y}\sin\frac{\pi k_x i}{N_x}\sin\frac{\pi k_y j}{N_y}\sin\frac{\pi k_x r}{N_x}\sin\frac{\pi k_y s}{N_y},
\end{split}
\end{align}    
where $\tilde{\Omega}_{k_x,k_y}=\Omega_{\frac{k_x}{2},\frac{k_y}{2}}$.
The rest of the method which we have reviewed briefly in section \ref{sec:Review} applies exactly in the same way in $(2+1)$-dimensions.

In this section we first work out entanglement entropy for a disk and a square entangling regions (see Fig.\ref{fig:shapes}). Our main concern here is to verify the emergence of $z$-dependent scaling behavior of entanglement measures (specifically EE). We also numerically verify that similar to Lorentzian theories, in Lifshitz field theories the singularities of the entangling regions bring in a logarithmic divergent term in EE.

The main difference between Lorentzian and Lifshitz cases is the corner contributions to EE are local in the former case but non-local in the latter one. It is well-known in the literature that corner contributions in Lorentzian theories are local effects and thus in $(2+1)$-dimensions which they have a logarithmic behavior, the coefficient of this logarithmic term is an additive function for any number of singularities in the entangling region. Here we present some arguments that the effect of dynamical exponents breaks down the latter property and as $z$ is increased the coefficient of the logarithmic term recedes from being additive.

\subsubsection*{Disk and Square Entangling Regions}

\begin{figure}
\centering
\includegraphics[scale=.43]{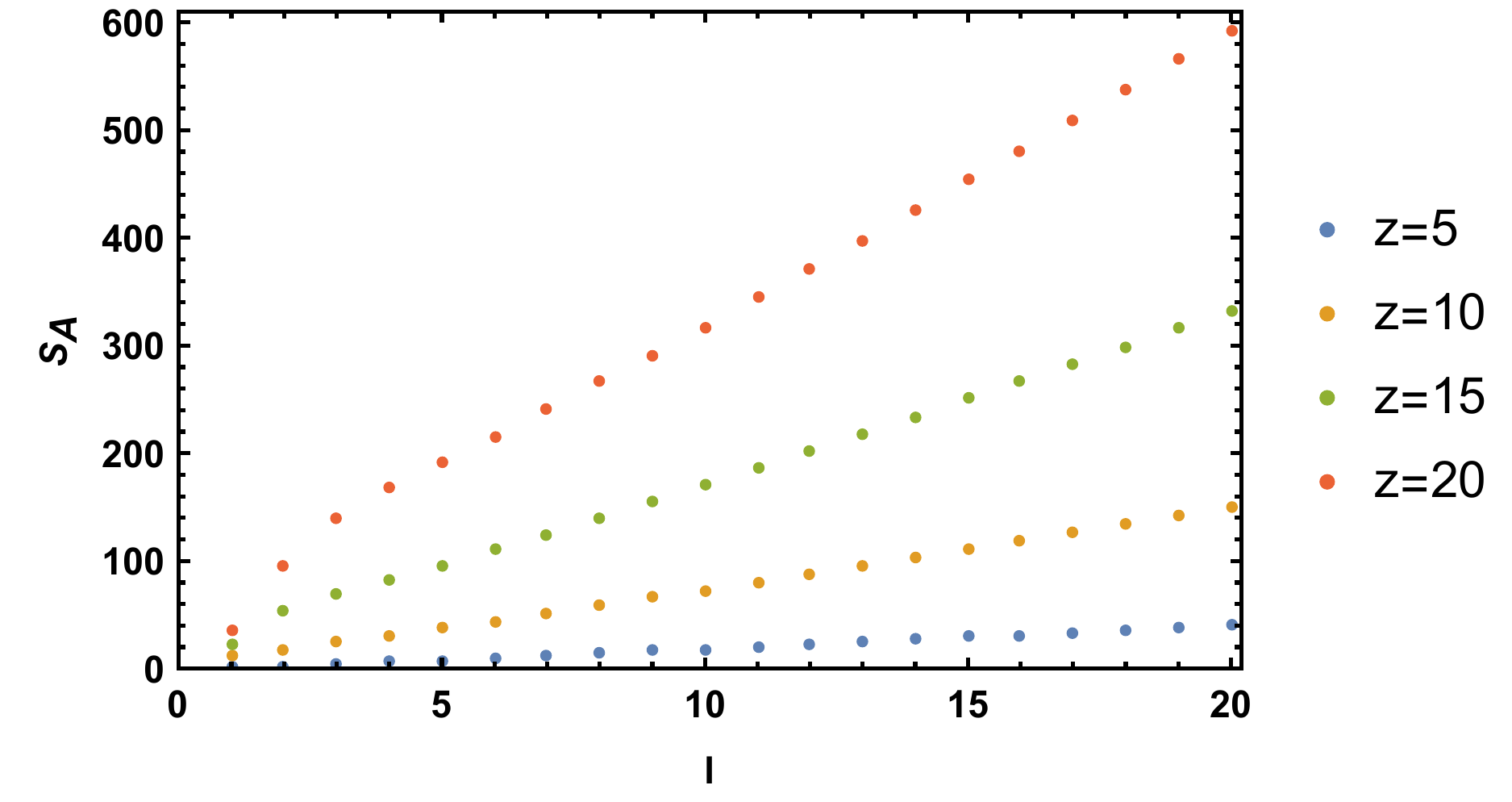}
\caption{The area law behavior of entanglement entropy for different values of $z$ for square entangling region. We have set $N_x=N_y=250$ and $m=0$ with Dirichlet boundary condition.}
\label{fig:P2p1SASq}
\end{figure}

\begin{figure}
\centering
\includegraphics[scale=.43]{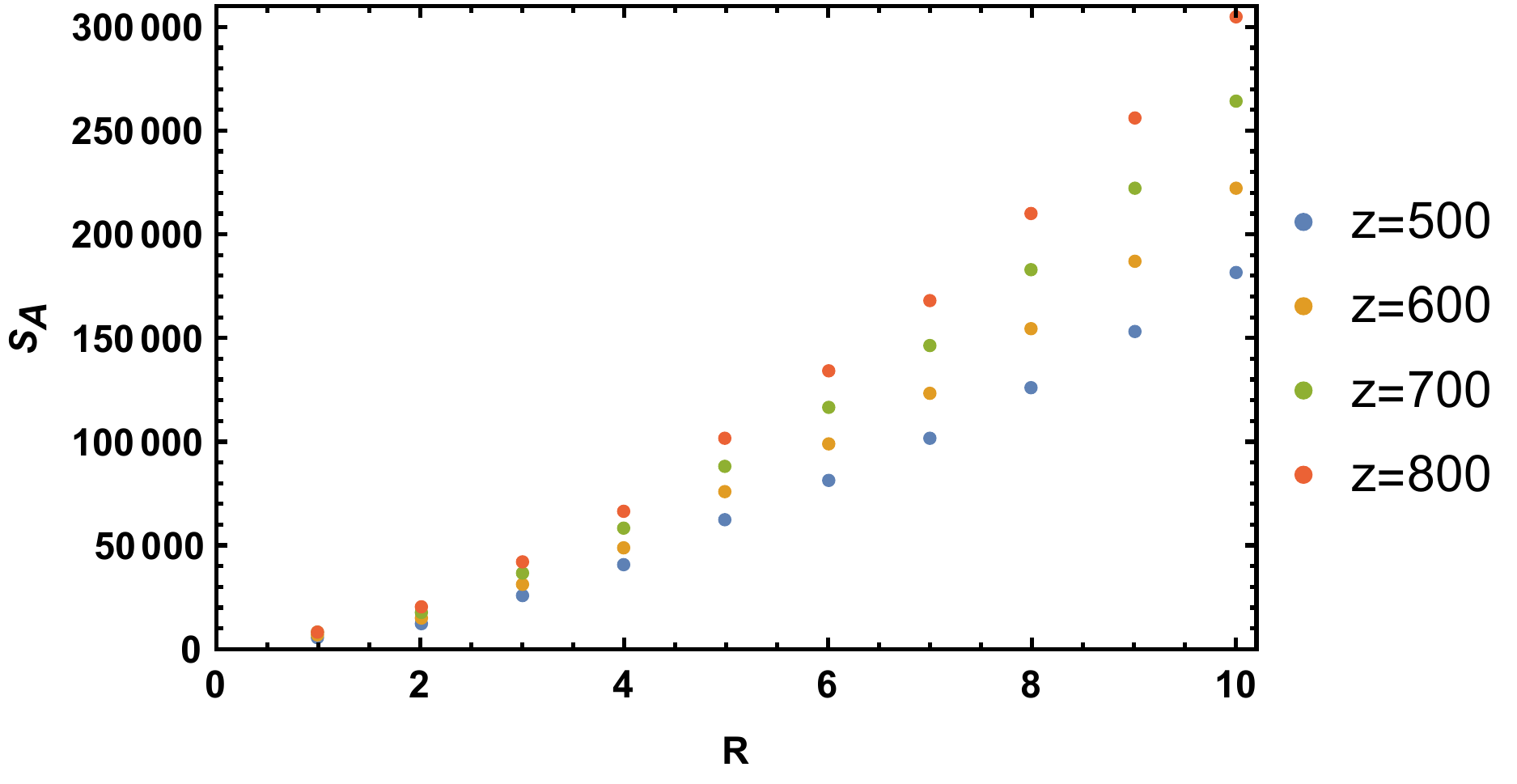}
\includegraphics[scale=.43]{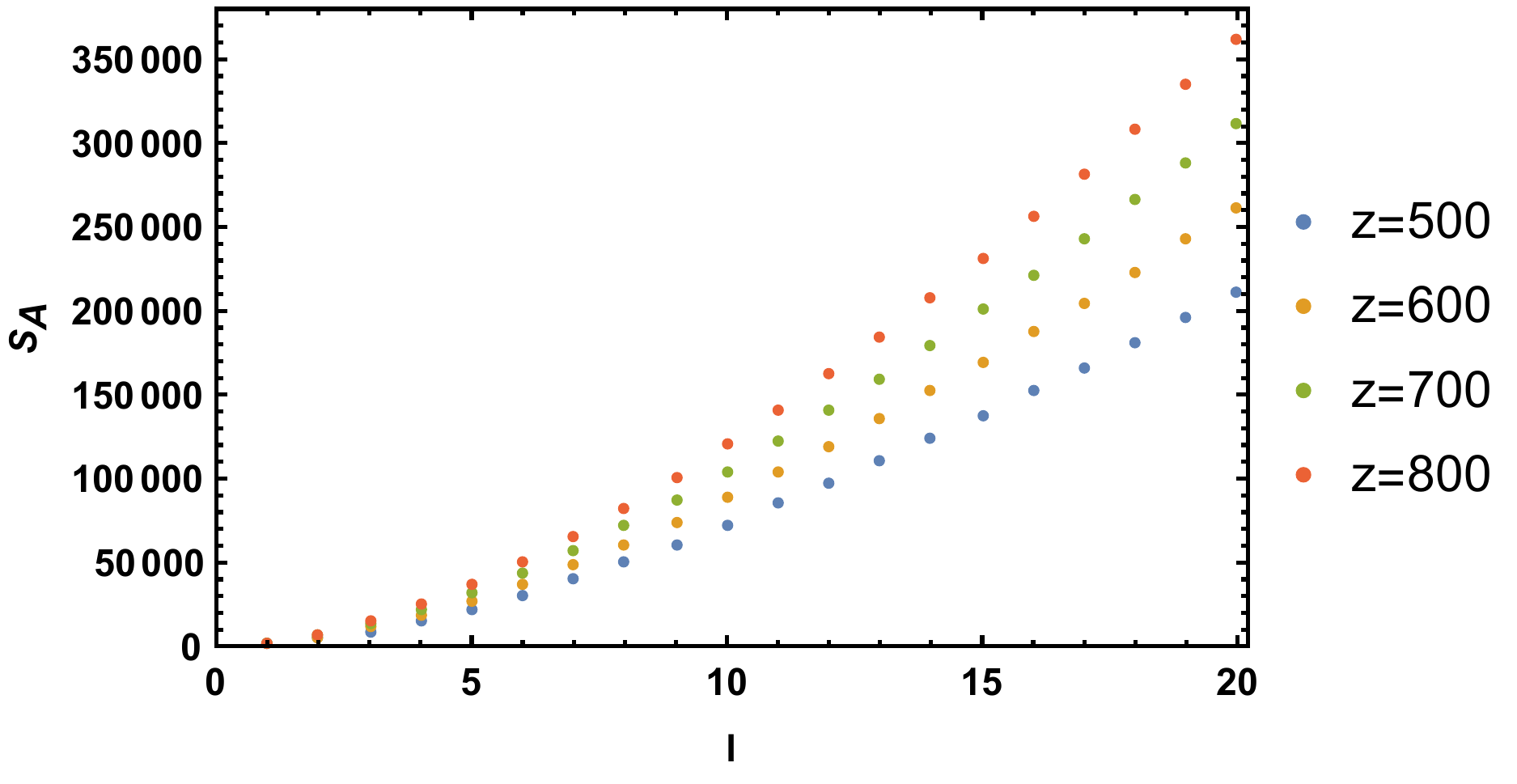}
\caption{Volume law for EE of a disk (with radius $R$) is shown in the left panel and a square (with side $\ell$) in the right panel in the large $z$ regime. We have set $N_x=N_y=250$ and $m=0$ with Dirichlet boundary condition.}
\label{fig:P2p1SA}
\end{figure}

In this subsection we focus on disk and square entangling regions (see the two left panels in Fig.\ref{fig:shapes}). Once again as we have explained previously (see Fig.\ref{fig:L-NL2p1}), due to the long tail of the kinetic term which causes $(2z+1)$ lattice sites to be correlated together, we expect that the EE is a monotonically increasing function of $z$. In Fig.\ref{fig:P2p1SASq} we have presented the area law behavior of entanglement entropy for square entangling region for $z\lesssim 20$. In Fig.\ref{fig:P2p1SA} we have shown the behavior of square and disk entangling regions for large values of $z$. One can see the change in the scaling of EE for regions inscribed in a square with length $20$. 

We are interested to investigate about the scaling of entanglement entropy for large enough dynamical exponent for different entangling regions. Our numerical data shows that the disk entangling region in $(2+1)$-dimensions for large enough $z$ and small enough regions fits with the following function 
\bea
S^{(z)}(\ell)=\mathcal{S}^{(z)}_{\rm vol.} \frac{\ell^2}{\epsilon^2}+\mathcal{S}^{(z)}_{\rm area} \frac{\ell}{\epsilon}+ S^{(z)}_0.
\eea
This function, up its first term which scales with the volume of the disk and is a sign of non-locality in this theory, is the same as what is well-known for disk entanglement in Lorentzian field theories. Our fitting data shows that similar to what we reported in the previous section for $(1+1)$-dimensions, for small regions here also the ratio of the volume term increases while the dynamical exponent increases.

There is one point we would like to clarify here about calculation of entanglement entropy of a disk in $(2+1)$-dimensions on a square lattice. It is sometimes mentioned in the literature that entanglement entropy is not a well-defined measure for a disk in $(2+1)$-dimensions since it is not a smooth function of the radius of the disk. This is because the EE of a disk shows significant fluctuates if one considers `continuous' radii for the disks. By continuous we mean radius values which are not multiples of the lattice constant. Due to these fluctuations people focus on mutual information as a more interesting measure for disk entangling region in $(2+1)$-dimensions \cite{Casini:2009sr}.\footnote{See also \cite{Casini:2015woa} where a $F$-theorem is introduced in $(2+1)$-dimensions using mutual information for co-centric disks.} Our numerical investigation shows that one can simply get rid of these fluctuations by focusing on disks which their radii is a multiple of the lattice spacing. As one can see there is no such fluctuation in our data in Fig.\ref{fig:P2p1SA}.

We have also calculated the entanglement entropy of a square entangling region. The main difference between square and disk entangling regions is the singularities (corners) present in the square. It is a well-known fact that these singularities  cause new divergent terms in entanglement entropy (see e.g. \cite{CasiniTalk}). Our numerical analysis shows that the entanglement entropy for small squares fits with the following function
\bea
S^{(z)}(\ell)=\mathcal{S}^{(z)}_{\rm vol.} \frac{\ell^2}{\epsilon^2}+\mathcal{S}^{(z)}_{\rm area} \frac{\ell}{\epsilon}+\mathcal{S}^{(z)}_{\rm corner} \log \frac{\ell}{\epsilon}+ S^{(z)}_0.
\eea 
We postpone a more careful study of the coefficient of this logarithmic term to the next subsection. Here we are interested to investigate how large is the contribution of the volume law to entanglement. Similar to what we reported for the disk, we found that the contribution of the volume law for small regions increases with the dynamical exponent $z$.

\begin{figure}
\centering
\includegraphics[scale=.37]{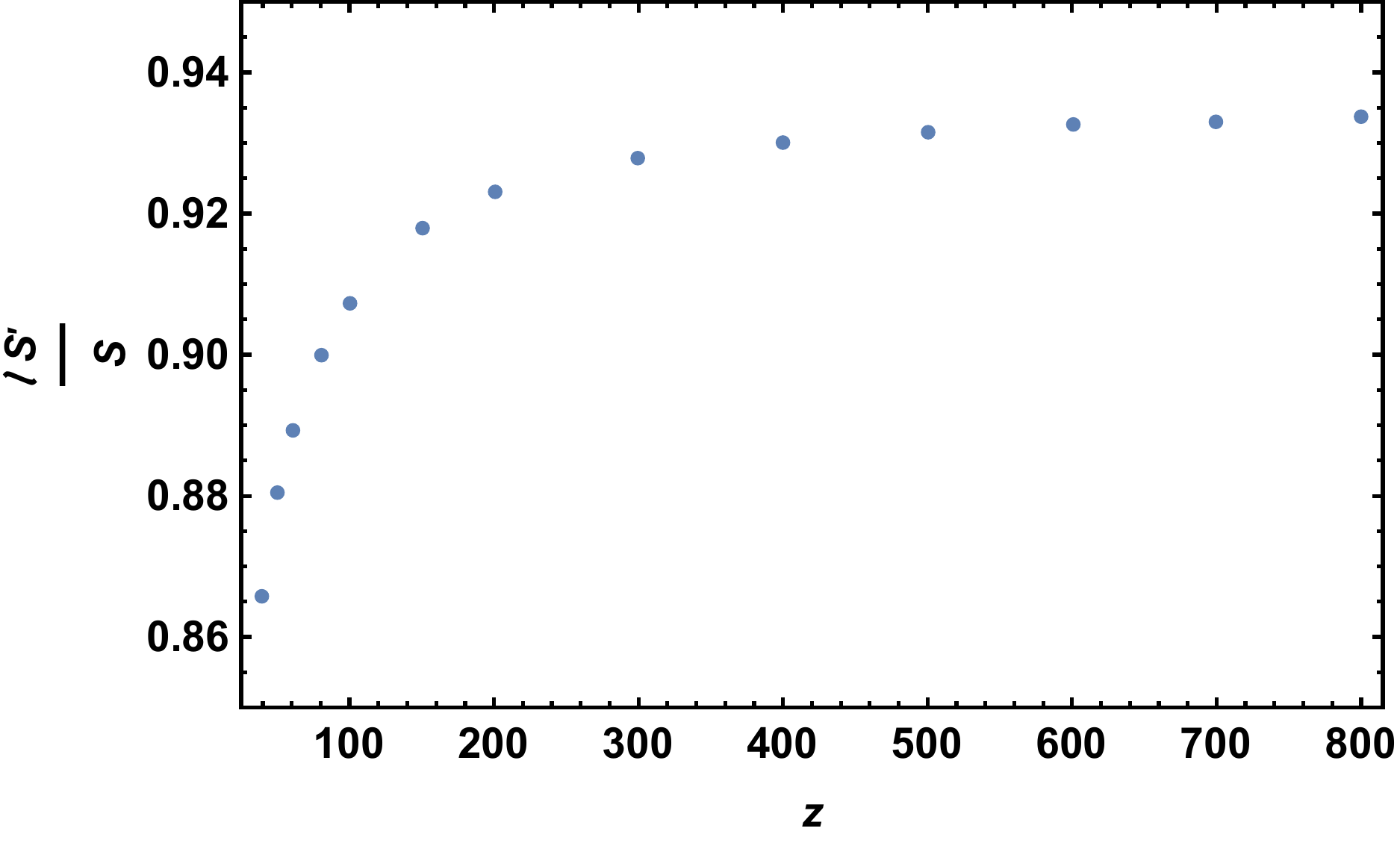}
\hspace{3mm}
\includegraphics[scale=.47]{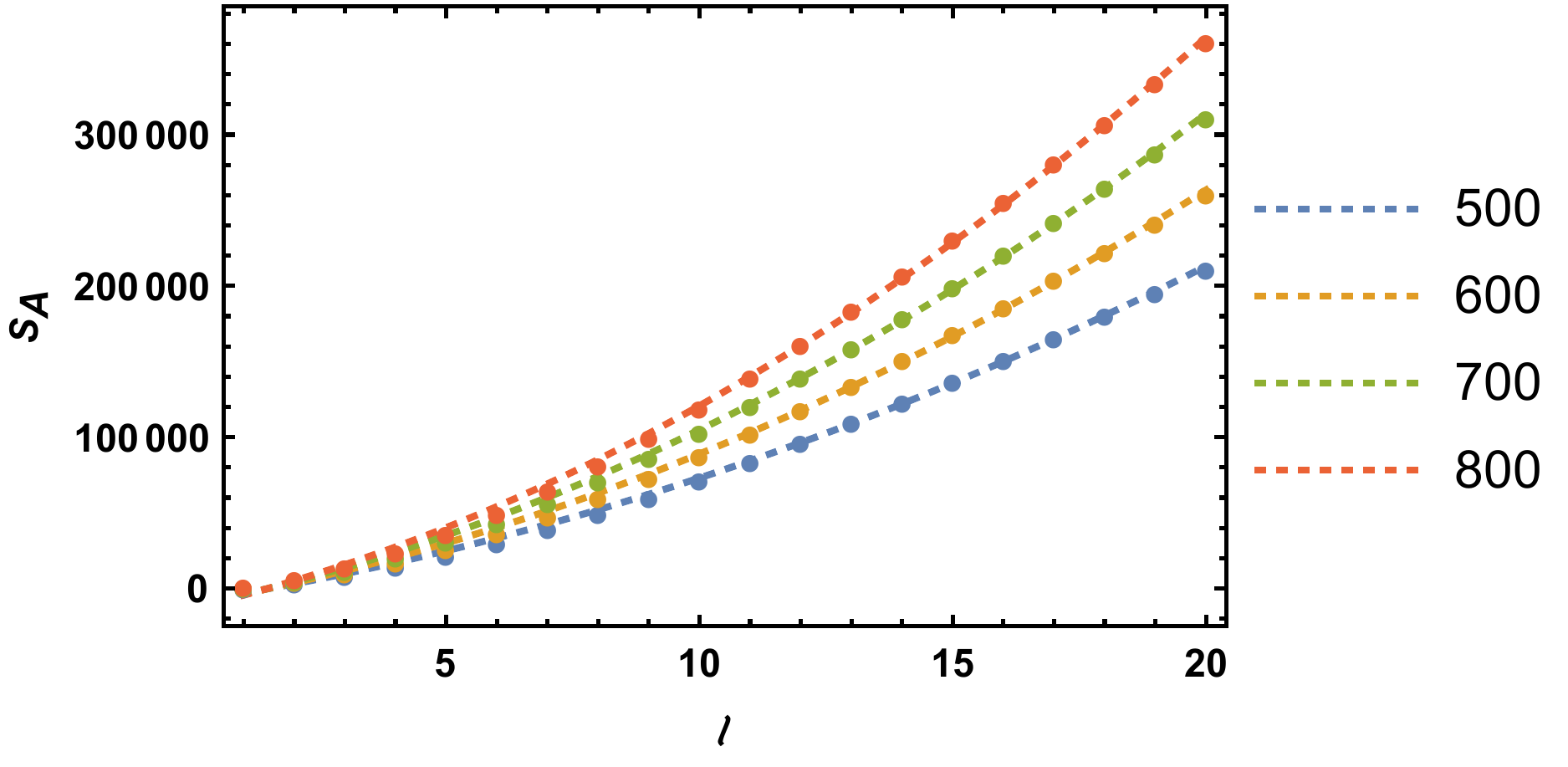}
\caption{\textit{Left}: $\frac{\ell S'}{S}$ as a function of $z$ for a square with enclosing 49 sites ($\ell=7$). For large values of $z$ this quantity approaches to a constant value exhibiting volume law behavior. \textit{right}: Fitting data for EE as a function of $\ell$ for different values of $z$. Here we consider Dirichlet boundary condition for a massless scalar with $N_x=N_y=250$.}
\label{fig:lsos}
\end{figure}

Fig.\ref{fig:lsos} is showing the same thing which we previously showed in Fig.\ref{fig:check} for (1+1)-dimensions. Here for a square entangling region with a fixed side $\ell$, we have plotted the $\frac{\ell  S'}{S}$ in the left panel. This quantity approaches a constant value which supports the volume law behavior in the $z\to\infty$ limit. In the right panel we have plotted the best fit for our proposed function for large values of $z$. 

\begin{figure}
\centering
\includegraphics[scale=.43]{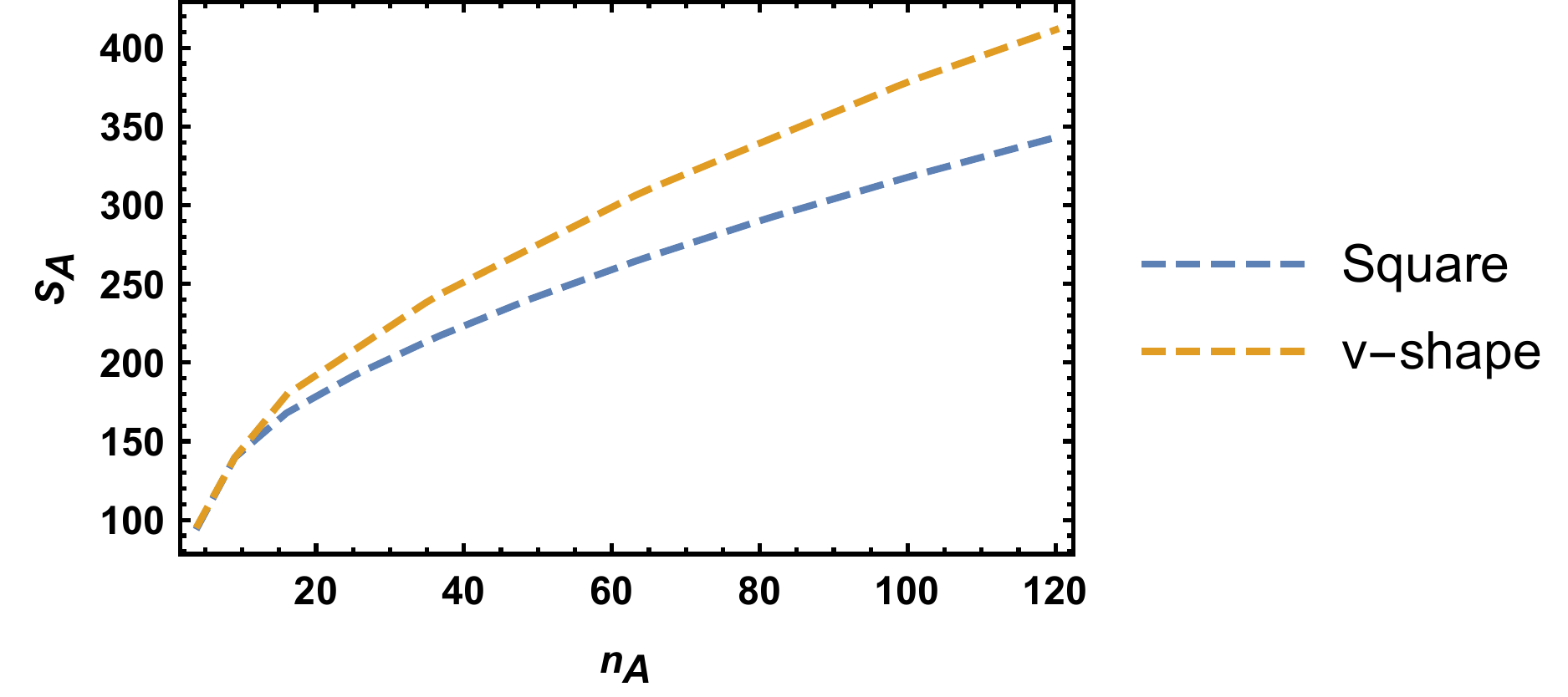}
\includegraphics[scale=.43]{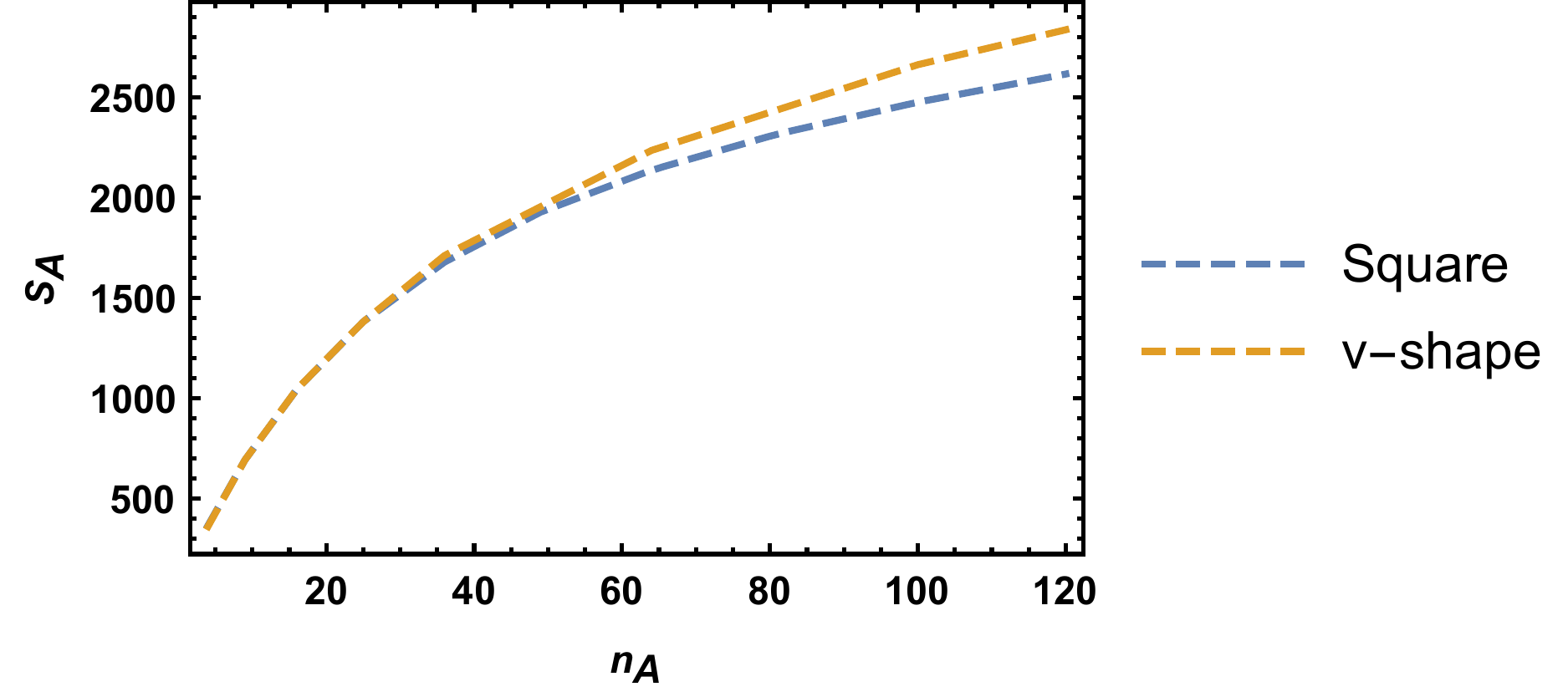}
\caption{Entanglement entropy as a function of number of sites enclosed by the entangling region, $n_A$, for square and v-shape. These two shapes have the same volume (see Fig.\ref{fig:shapes}). In the left panel we have set $z=20$ and in the right panel $z=50$. For small regions, where we expect a volume law, the entropy for both shapes coincides. As $n_A$ increases, the entanglement of the v-shape overtakes the square since it has a larger area. We have set $N_x=N_y=250$ and $m=0$ with Dirichlet boundary condition.}
\label{fig:P2p1VE}
\end{figure}

In Fig.\ref{fig:P2p1VE} we have plotted the entropy as function of the volume, i.e. number of sites enclosed by the entangling region. This is done for square and v-shape which have the same volume. One can find that for small regions where the entropy scales with the volume the entropy of these two shapes coincides but for larger regions the entropy of the region with a larger area grows faster. For larger values of $z$ the region of coincidence increases. 

\begin{figure}
\begin{center}
\includegraphics[scale=0.5]{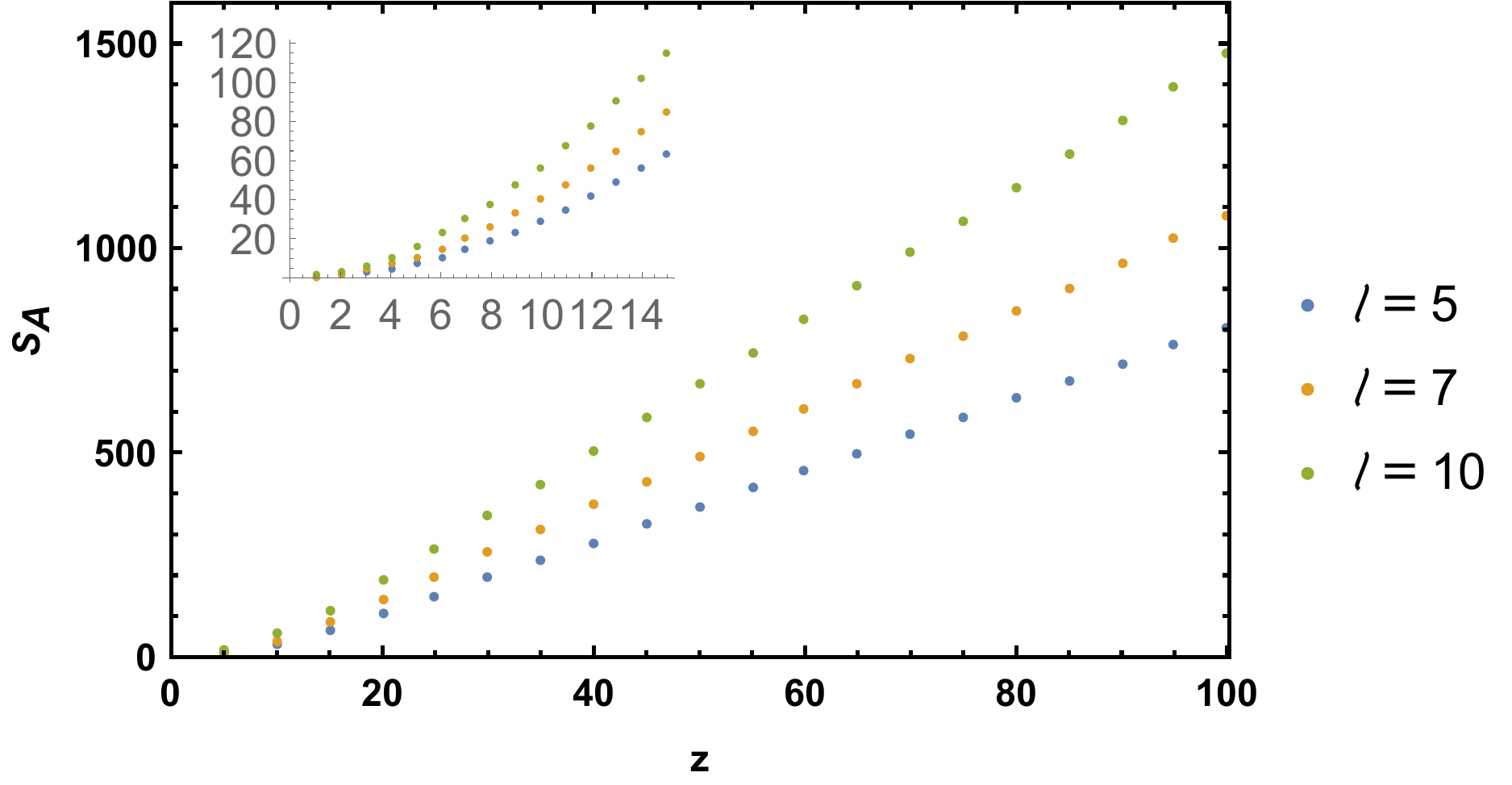}
\end{center}
\caption{EE as a function of $z$ for square entangling regions with different size. The internal panel focuses on $1<z<15$. We have set $N_x=N_y=250$ and $m=0$ with Dirichlet boundary condition.}
\label{fig:Sofz2}
\end{figure}

We have also studied the $z$-dependence of entanglement entropy. In Fig.\ref{fig:Sofz2} we have shown the results for square entangling regions with different size. Our results is similar to what we have found in (1+1)-dimensions. As the dynamical exponent increases from the Lorentzian case, for small values of $z$ the entropy grows quadratically which the length of validity for this regime depends on the size of the entangling region. After this regime the entropy growth enters a linear regime which our results does not show any stop for it.

Our results show that for generic dynamical exponent $z$, we expect the leading term of entanglement entropy (and also Renyi entropies) to be a function which interplay between these two behaviors. The simplest choice with this property in $(2+1)$-dimensions is   	
$$\left(\frac{\ell}{\epsilon}\right)^{2-\frac{1}{z}}.$$

\subsubsection*{Corner Contributions to EE}
Up to now we have shown different features of the intrinsic non-locality due to the dynamical exponent in the entanglement structure of our Lifshitz theory of interest. In this subsection we will continue our study focusing on the corner contributions to entanglement entropy in $(2+1)$-dimensions. Corner contributions to entanglement entropy are well-known to be local effects. In a local $(2+1)$-dimensional theory corner contributions appear as subleading logarithmic divergent term \cite{Casini:2006hu, Casini:2008as, Casini:2009sr}
\be
S_A=\mathcal{S}_{\rm area} \frac{\ell}{\epsilon}+\mathcal{S}_{\rm corner} \log \frac{\ell}{\epsilon}+ S_0.
\ee  
where
$$\mathcal{S}_{\rm corner}=\sum_{\rm{corners\,on}\,\partial A}a\left(\theta_i\right),$$
and $\theta_i$ is the opening angle of the $i$-th corner. Since the corner contribution is a universal term in the entanglement entropy expansion, a very important question is what kind of characteristic information of the theory it contains. Although we do not know the answer to this question in general, recently in the context of AdS/CFT\footnote{See \cite{Hirata:2006jx, Myers:2012vs} for detailed study of corner contributions in this context.}, it has been shown that in the smooth limit, i.e. $\theta\to\pi$, the coefficient of this term is the same as what appears in the two point function of the stress tensor of the underlying conformal field theory \cite{Bueno:2015rda}.

In order to study these singularity effects in a Lorentzian theory one can consider entangling regions with the same area but different number of singularities. See the two middle panels of Fig.\ref{fig:shapes} which show a square (with four corners) and the a-shape which is a deformation of the square which preserves the area but has six corners. In this case we can forget about the leading area term and focus on the subleading logarithmic term which is supposed to capture the difference between these two regions. For these two regions if one works out the entanglement entropy and find the logarithmic term, we expect (from locality) to find
$$\mathcal{S}^{({\rm a-shape})}_{\rm corner}=\frac{3}{2}\, \mathcal{S}^{({\rm square})}_{\rm corner},$$  
which can be verified numerically (see e.g. \cite{CasiniTalk}).

In the case of non-local theories, one would consider two regions with the same volume but different number of corners. In the left panel of Fig.\ref{fig:shapes} we have shown a deformed square which has the same volume as the square but eight corners (we refer to this shape with index `v'). Our numerical results for both `a' and `v' entangling regions show that as the dynamical exponent increases the ratio of the corner contribution to that of a square starts to decrease which shows that corner contributions are no more local effects.

\begin{figure}
\centering
\includegraphics[scale=.3]{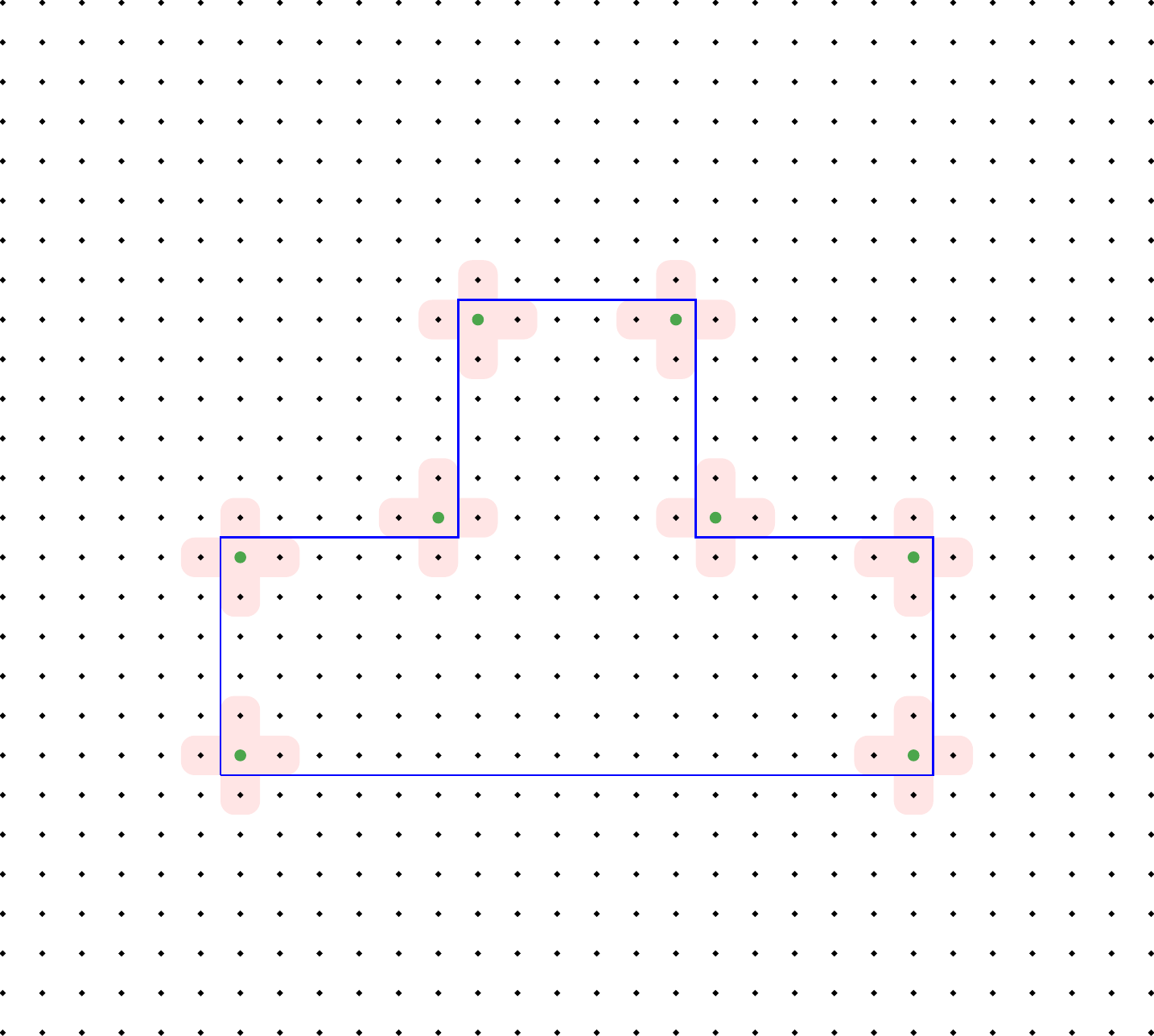}
\hspace{3mm}
\includegraphics[scale=.3]{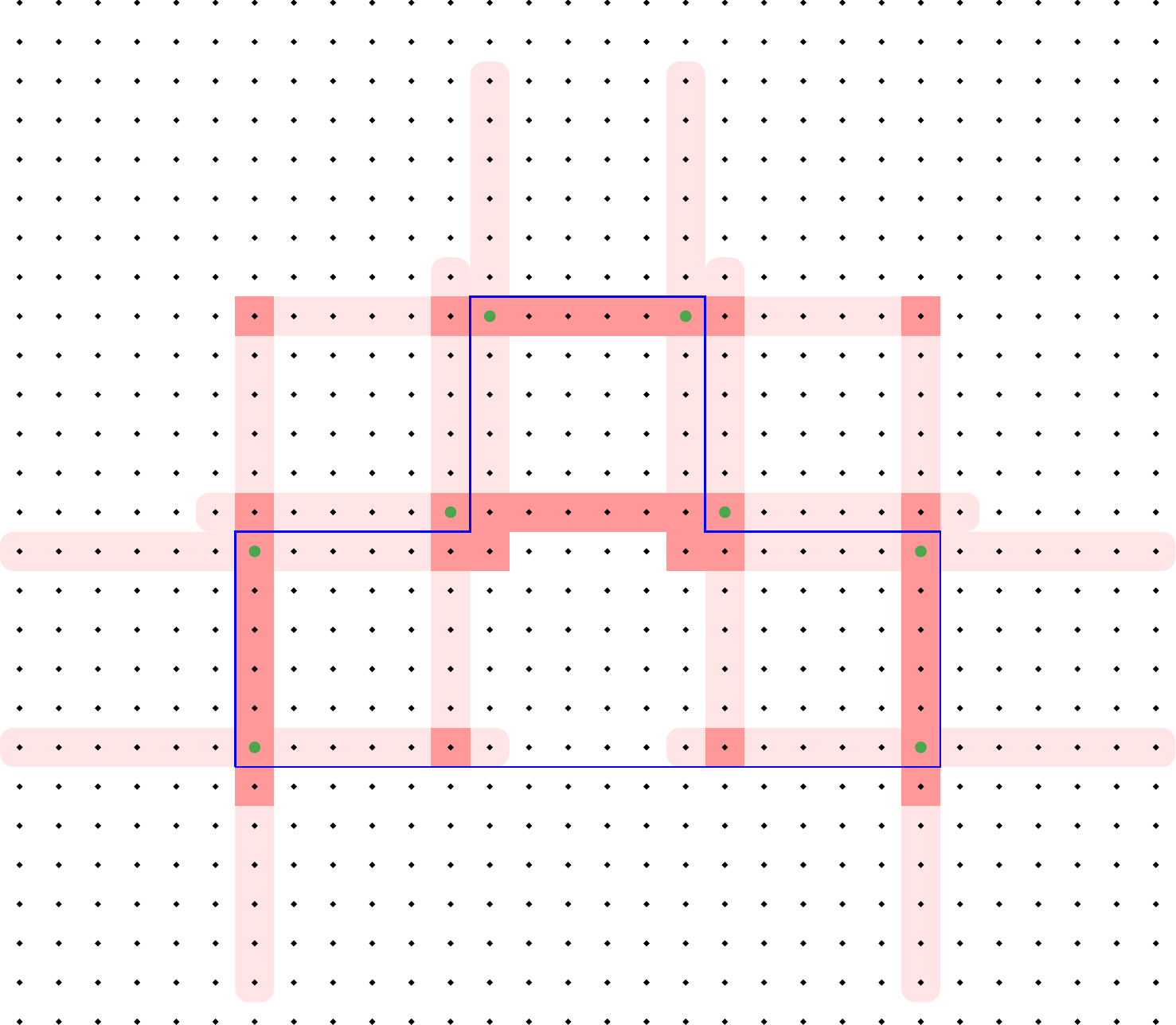}
\caption{Illustration of mixing of corner contributions in Lifshitz theories. The left belongs to $z=1$ case which there is no mixing and the logarithmic term in the EE is additive. The right panel belongs to $z=6$ which there are several points which are correlated to more than one green point, thus the corner contributions are no more additive.}
\label{fig:P2p1locality2}
\end{figure}

The way we can explain what happens in Lifshitz theory is as follows: contribution of each corner is mixed up with others due to correlation of $(2z+1)$ lattice cites. This has been demonstrated in Fig.\ref{fig:P2p1locality2}. In order to have a neat picture we have just focused on the nearest points to the singularity which are show in green. This is done for `v'-shape and the similar thing happens for the `a'-shape region. The red shadowed points are those which are in the $(2z+1)$ neighbourhood of the green points. The left panel belongs to the Lorentzian case $z=1$ which non of the shadowed points of each green point coincide with each other. The right panel belongs to $z=6$. In this case there are lots of shadowed points which belong to the shadow of more than one corner (green) point. The number of such points increases with the dynamical exponent and this is why we think the corner contributions are no more additive in the case of Lifshitz scalar theory. This mixing procedure will also happen in Lorentzian theories if there are adjacent corners in the entangling region \cite{Mozaffar:2015xue}.

\subsection{Entanglement Measures in Massive Scalar Theory}\label{sec:massive}
In this section we continue our study by considering the massive scalar field theory. In order to investigate the effects of nonzero mass on the entanglement we will consider the periodic boundary condition for the field in different dimensions. The results for the entanglement and Renyi entropy in ($1+1$)-dimensions are summarized in Fig.\ref{fig:EE-periodic}. 

According to these results similar to the massless case the EE increases with the dynamical exponent. The important thing to note here is that the massive case is completely different from the massless case. We have not found volume law regime for small subregions at any value of $z$ in our study. Our results shows that for periodic boundary conditions, for mass parameters which lead to stable numerical results, we always recover an area law scaling for entanglement entropy. Increasing the mass parameter makes the entanglement to grow slower as a function of the subregion length and also the dynamical exponent. We have plotted some results for entanglement and Renyi entropies in Fig.\ref{fig:EE-periodic}.

In case of Dirichlet boundary condition for small mass parameters in the $m\to0$ limit we recover the results reported in the massless section. We have not found any range of parameters even with large dynamical exponents with a volume law behavior for small regions. This is in agreement with the behavior of the correlator in ($1+1$)-dimensions in the small $r$ limit which is given by
\be
\langle\phi(0)\phi(r)\rangle = \frac{\sec\left(\frac{\pi}{2z}\right)}{2\sqrt{2}\pi }\frac{\Gamma\left(\frac{2z+1}{2z}\right)}{\Gamma\left(\frac{z+1}{2z}\right)}\frac{1}{m^{z-1}}-\frac{1}{8\sqrt{2}z\pi^2 }\Gamma\left(\frac{3}{2z}\right)\Gamma\left(\frac{z-3}{2z}\right)\frac{r^2}{m^{z-3}}+\mathcal{O}\left(r^3\right),
\ee
for integer dynamical exponent $z>3$.
The scaling of the leading term does not depend on the dynamical exponent and is divergent in the $m\to0$ limit.
This causes our intuitive picture (illustrated in Fig.\ref{fig:L-NL2p1}) not to be valid in the massive regime. A more careful analysis is needed to understand the massive regime of this theory which we postpone it for future works. 

\section{Conclusions and Discussions}
We have mainly studied the entanglement structure of a free massless scalar theory (see equation \eqref{action}) with Lifshitz scaling symmetry. This family of theories are supposed to describe systems at quantum critical points. We have analysed our theory of interest in 1+1 and 2+1 space-time dimensions. We explored the role of the dynamical exponent in entanglement measures. We would like to first summarize our main results:

\begin{figure}
\centering
\includegraphics[scale=.75]{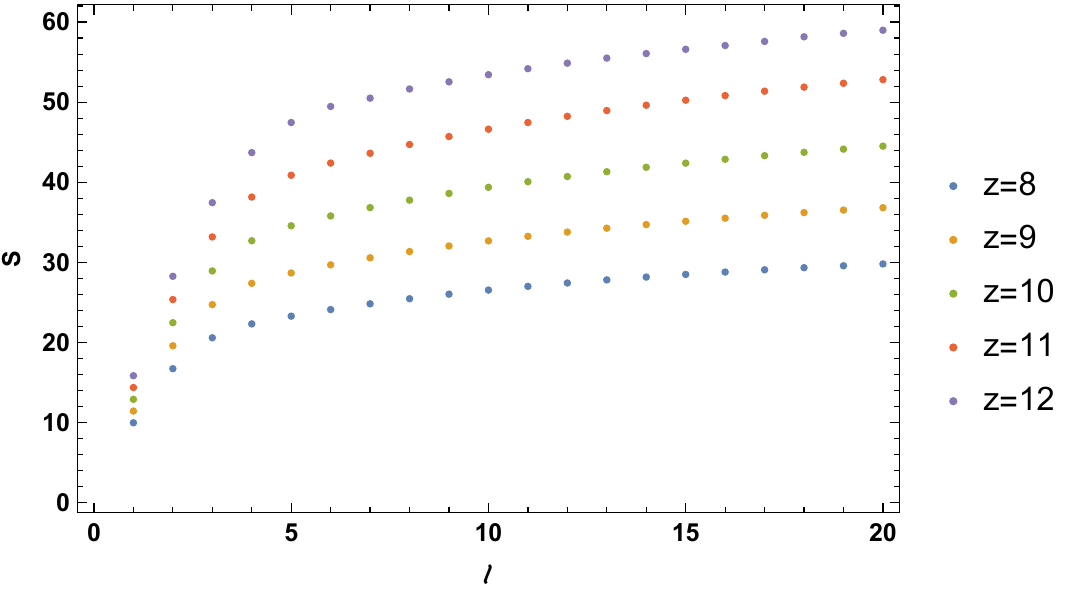}
\includegraphics[scale=.75]{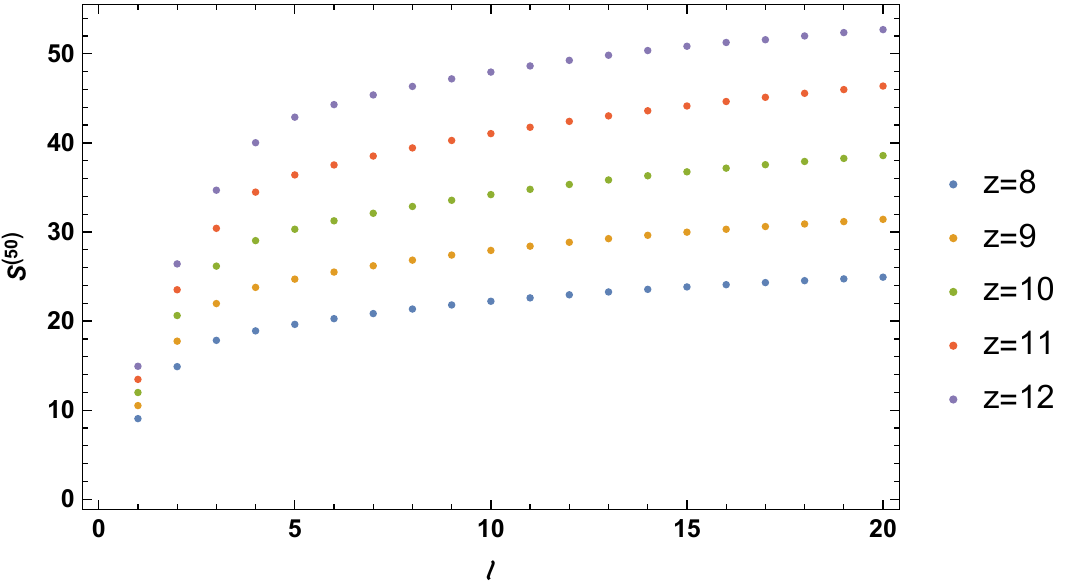}
\caption{Entanglement and Renyi entropy for periodic boundary condition in $1+1$ dimensions with $m=0.1$ and $N=1000$.}
\label{fig:EE-periodic}
\end{figure}

\begin{itemize}
\item In the massless limit for small entangling regions the leading part in the EE expansion transits from area law to volume law for large dynamical exponents. Regarding to our results in 1 and 2 spatial dimensions for several entangling regions, we propose the following scaling behavior as the leading term of EE in a $(d+1)$-dimensional theory
\bea\label{scalingEE}
S^{(z)}(\ell)=\#\left(\frac{\ell}{\epsilon}\right)^{d-\frac{1}{z}}+\cdots,
\eea
where $\ell$ is a characteristic length of the entangling region.
Indeed, the spatial part of the kinetic term in this theory couples $(2z+1)$ lattice sites together at each point, and therefore the value of entanglement grows due to the correlation between more points in comparison with the Lorentzian case. In other words the theory shows more non-local effects for larger dynamical exponents. This is why one would naturally expect a volume law scaling rather than an area law for small entangling regions. Regarding to this result we would like to also note that this result is in contrast with what has been previously reported in \cite{Solodukhin:2009sk} where the author has argued that in any theory with a generic form of dispersion relation as $\omega=f(k^2)$, where $k^\mu$ is the momentum vector, the EE scales with the area of the entangling region.

\item We have shown that the mutual information is an increasing function of the dynamical exponent $z$. Also the fall off mutual information in large separation limit is slower in comparing to the relativistic case. This behavior is also due to stronger correlations between lattice points for larger values of $z$. 

\item Similar to free scalar Lorentzian theory, the resultant tripartite information is always positive for all values of $z$. This shows that at least the vacuum state of this theory can not have a solution of a classical gravity theory as a holographic dual. This is the main clue that forbids us to compare our results (especially Eq.\eqref{scalingEE}) with some previous studies on Lifshitz theories in the context of gauge/gravity duality. 

\item In 2 spatial dimensions we have argued that corner contributes which are known to be local effects in Lorentzian field theories become more and more non-local while the dynamical exponent increases. As a result corner contribution to entanglement entropy are no more additive in these theories. 
\end{itemize}

We have also observed that for massive theory the behavior of the $m\to0$ limit at fixed $z$ is very similar to the behavior of large $z$ at fixed $m$. Although at the moment it is not an easy task to understand this behavior from the correlation functions for generic $z$ because of technical reasons, but it is worth to note that this behavior is in agreement with what is known about a massive scalar field propagating on a Lifshitz background in the context of holography (see Eq. 3.23 of \cite{Kachru:2008yh}). 

Of course it would be of great interest if one can verify our results with a concrete analytical analysis. As an example one can study propagators of free scalar theories on a $n$-sheeted space-time and work out the leading behavior entanglement and Renyi entropies analytically. Recently a nice paper appeared where heat kernels for non-relativistic field theories on specific manifolds has been worked out in it \cite{Pal:2017ntk, Barvinsky:2017mal}. In principle it is possible to extend this study to singular manifolds and work out the entanglement entropy.  

Finally we would like to note some interesting questions which generate new directions for our analysis.
One way to extend our study is to go beyond the vacuum state. In the case of mixed state, such as thermal states it is interesting to analyse other entanglement measures which are more suitable for mixed state, e.g., logarithmic negativity \cite{Calabrese:2012ew,Eisler2015,DeNobili:2016nmj}.

Another interesting question is about the time evolution of EE in such theories which is now a typical way to study universal behaviors of these family of theories in out of equilibrium phases. One can study entanglement measures under quantum quenches and study the corresponding thermalization process. Indeed, this analysis may shed light on the role of critical exponent on the entanglement structure. We will report some interesting results in these directions in early future \cite{inprog}.

\section*{Acknowledgements}
We would like to thank Cristiano De Nobili, Viktor Eisler, Willy Fischler, Igor Peschel, Charles Rabideau, Sergey Solodukhin, Yasaman Yazdi and specially Andrea Coser and Noburo Shiba for valuable correspondence on related subjects to this work.
We are very grateful to Mohsen Alishahiha, Mohammad Ali Rajabpour, Shahin Sheikh-Jabbari and Tadashi Takayanagi for their useful comments on an early draft of this work.
We also thank Amin Akhavan, Amin Faraji Astaneh, Ali Naseh, M. Reza Tanhayi and Mohammad H. Vahidinia for related discussions.
MRMM would like to thank CERN-TH Division for their warm hospitality during parts of this work.


\end{document}